\documentclass[fleqn,usenatbib]{mnras} 

\usepackage{newtxtext,newtxmath}
\usepackage[T1]{fontenc}

\DeclareRobustCommand{\VAN}[3]{#2}
\let\VANthebibliography\thebibliography
\def\thebibliography{\DeclareRobustCommand{\VAN}[3]{##3}\VANthebibliography}

\usepackage{graphicx}
\usepackage{amsmath}
\usepackage{siunitx}
\usepackage[capitalize]{cleveref}
\usepackage{multirow}
\usepackage{threeparttable} 
\usepackage{nth}

\newcommand{\eg}{e.g.,}
\newcommand{\ie}{i.e.,}
\newcommand{\qsi}{\ensuremath{q_\mathrm{sil}}}
\newcommand{\qcarb}{\ensuremath{q_\mathrm{carb}}}
\newcommand{\qwa}{\ensuremath{q_\mathrm{ice}}}
\newcommand{\qwatot}{\ensuremath{\phi_\mathrm{ice}}}
\newcommand{\por}{\ensuremath{p}}
\newcommand{\portot}{\ensuremath{\phi_\mathrm{vac}}}
\newcommand{\qdnorm}{\ensuremath{Q_\mathrm{0}}}
\newcommand{\qdslope}{\ensuremath{b_\mathrm{s}}}
\newcommand{\qdstar}{\ensuremath{Q_\mathrm{D}^\star}}
\newcommand{\qdstarcon}{\ensuremath{Q_\mathrm{D,\,\qty{32}{\um}}^\star}}
\newcommand{\etacon}{\ensuremath{\eta_{0,\,\qty{32}{\um}}}}

\newcommand{\chitot}{\ensuremath{\widehat{\chi^{\raisebox{-.5ex}{$\scriptscriptstyle 2$}}}}}
\newcommand{\chirad}{\ensuremath{\chi_\mathrm{rad}^2}}
\newcommand{\chised}{\ensuremath{\chi_\mathrm{SED}^2}}
\newcommand{\diambl}{\ensuremath{D_{\mathrm{bl}}}}
\newcommand{\rave}{\texttt{RAVE}}
\newcommand{\webbpsf}{\texttt{WebbPSF}}
\newcommand{\tinytim}{\texttt{STinyTim}}
\newcommand{\mpl}{\ensuremath{M_\mathrm{pl}}}
\newcommand{\apl}{\ensuremath{a_\mathrm{pl}}}

\DeclareSIUnit{\year}{yr}
\DeclareSIUnit{\jansky}{Jy}
\DeclareSIUnit{\arcsecword}{arcsec}
\DeclareSIUnit{\erg}{erg}
\DeclareSIUnit{\milliarcsec}{mas}
\DeclareSIUnit\jupitmass{\ensuremath{\mathit{M}_{\mathrm{Jupiter}}}}
\DeclareSIUnit\saturmass{\ensuremath{\mathit{M}_{\mathrm{Saturn}}}}
\DeclareSIUnit\solarmass{\ensuremath{\mathit{M}_{\odot}}}
\DeclareSIUnit\solarlum{\ensuremath{\mathit{L}_{\odot}}}
\title[A PR drag model for the Fomalhaut disk]{
    A PR drag origin for the Fomalhaut disk's pervasive inner dust: constraints on collisional strengths, icy composition, and embedded planets}

\author[M.~Sommer et al.]{
    Max Sommer,$^{1}$\thanks{E-mail: ms3078@cam.ac.uk}
    Mark Wyatt,$^{1}$
    and Yinuo Han$^{1,2}$
    \\
    $^{1}$Institute of Astronomy, University of Cambridge, Madingley Road, Cambridge, CB3 0HA, UK\\
    $^{2}$Division of Geological and Planetary Sciences, California Institute of Technology, 1200 E. California Blvd., Pasadena, CA 91125, USA\\
    }

\date{Accepted 2025 March 19. Received 2025 February 27; in original form 2024 December 12}

\pubyear{2025}

\begin{document} 

  \label{firstpage}
  \pagerange{\pageref{firstpage}--\pageref{lastpage}}
  \maketitle
  \begin{abstract}
    Recent JWST observations of the Fomalhaut debris disk have revealed a significant abundance
    of dust interior to the outer planetesimal belt, 
    raising questions about its origin and maintenance. 
    In this study, we apply an analytical model to the Fomalhaut system,
    that simulates the dust distribution interior to a planetesimal belt, 
    as collisional fragments across a range of sizes are dragged inward under Poynting-Robertson (PR) drag.
    We generate spectral energy distributions and synthetic JWST/MIRI images of the model disks,
    and perform an extensive grid search for particle parameters---pertaining to 
    composition and collisional strength---that best match the observations.
    We find that a sound fit can be found for particle properties
    that involve a substantial water ice component, around 50\%--80\% by total volume,
    and a catastrophic disruption threshold, \qdstar{}, at a particle size of 
    $D\!\approx\!\qty{30}{\um}$ of \mbox{(\numrange{2}{4})$\,\times\qty{e6}{\erg\per\g}$}.
    Based on the expected dynamical depletion of migrating dust by an intervening planet 
    we discount planets with masses 
    \qty{>1}{\saturmass} beyond \qty{\sim50}{\astronomicalunit} in the extended disk,
    though a planet shepherding the inner edge of the outer belt of up to \qty{\sim2}{\saturmass}
    is reconcilable with the PR-drag-maintained disk scenario, contingent upon higher collisional strengths.
    These results indicate that PR drag transport from the outer belt alone can account
    for the high interior dust contents seen in the Fomalhaut system, 
    which may thus constitute a common phenomenon in other belt-bearing systems.
    This establishes a framework for interpreting mid-planetary system dust
    around other stars, with our results for Fomalhaut providing a valuable calibration of the models.
\end{abstract}


  \section{Introduction}
The Fomalhaut system is known as one of the archetype debris disks owing to its 
luminosity and proximity (\qty{7.7}{pc}).
Following the first detection of the disk's substantial infrared excess emission
by the Infrared Astronomical Satellite (IRAS) \citep{Gillett1986iras}, 
Fomalhaut's disk garnered significant attention
due to its apparent multi-component structure.
Observations across the electromagnetic spectrum---from
scattered light captured by \textit{Hubble} \citep{Kalas2005planetary,Kalas2013stis},
to the infrared with \textit{Spitzer} and \textit{Herschel} 
\citep{Stapelfeldt2004first,Acke2012herschel},
and at submillimetre/millimetre wavelengths with the James Clerk Maxwell Telescope (JCMT)
and the Atacama Large Millimeter Array (ALMA) \citep{Holland2003submm,MacGregor2017complete}, 
amongst others---indicated an intricate system,
including an outer belt at \qty{\sim140}{\astronomicalunit},
as well as an inner, warm (\qty{\sim150}{\kelvin}) disk component
localized at \qtyrange{8}{15}{\astronomicalunit}, 
as inferred from the spectral energy distribution (SED) 
of the total emission \citep{Su2016inner}.
This configuration seemed to resemble the planetesimal reservoirs of the solar system,
with a cold Kuiper belt and a warmer, relatively narrow asteroid belt.

However, recently, the James Webb Space Telescope (JWST) changed this conception
of the Fomalhaut system, which for the first time resolved the distribution
of dust interior to the outer belt \citep{Gaspar2023spatially}.
These observations reveal a prominent drawn-out inner disk 
from around \qtyrange{10}{90}{\astronomicalunit},
in addition to an `intermediate belt', an eccentric ($e\!\approx\!0.3$) ring-shaped
brightness feature stretching from roughly \qtyrange{60}{110}{\astronomicalunit},
which together fill up much of the space interior to the outer belt.

Such a continuous distribution of dust permeating the space interior to
the outer planetesimal belt raises questions about its origin, in particular,
whether it can be maintained by dust produced in the outer belt
and being dragged inward by Poynting-Robertson (PR) drag, 
or whether other scenarios, such as dust delivery by comets 
\citep[\eg][]{Faramaz2017inner,Marino2017alma}, are necessary.
This also pertains to the intermediate belt, which could conceivably be caused 
by a planet trapping migrating dust in orbital resonances, or 
by \textit{in situ} dust production in a spatially confined, 
collisionally grinding planetesimal reservoir.

\citet{Wyatt2005insignificance} provided an idealized approach to gauge the abundance of 
PR-drag-supplied dust interior to a cold belt---assuming 
disks to be composed of only the smallest, barely-bound grains---which
indicated the mechanism's ineffectiveness to produce signals detectable 
with the contemporary instrumentation.
Then, in light of more sensitive interferometric observing methods becoming available,
\citet{Kennedy2015warm} concluded that PR drag transport could no longer be 
considered insignificant, using a similar approach.
Recent advancements of these analytical methods to incorporate
a distribution of particle sizes \citep{Rigley2020dust},
now allow a more rigorous reassessment of the PR drag scenario.
In this study, we adapt and apply the model developed by \citet{Rigley2020dust}
to the Fomalhaut disk, to evaluate whether a pure PR drag transport scenario
can account for the newly observed characteristics of the extended disk.
Specifically, we simulate disks with a broad range of particle material parameters---facilitated
by the fast analytical model---seeking outcomes consistent with both the disk's SED,
and the radial brightness profiles extracted from JWST images. 
In doing so, we incorporate more definitive properties of the outer belt 
derived from ALMA imaging as fixed model inputs. 
Given the simplicity of the model, the aim of this study is to assess 
the general validity of a PR-drag-fed extended disk,
and does not target the reproduction of more intricate features,
such as the eccentricity of the outer belt or the intermediate ring,
or the enigmatic near-infrared excesses detected in interferometric observations 
attributed to a population of close-in hot dust 
\citep{Mennesson2013interferometric,Lebreton2013interferometric,Su2016inner}

\cref{Sect:Obs} provides an overview of the observational data.
\cref{Sect:Method} introduces the PR drag disk model and details
our method to generate and evaluate synthetic observations,
across a grid of model parameters.
\cref{Sect:Fitting} analyses the parameter dependence of suitable model outcomes.
\cref{sect:best_fit} presents the best-fit model outcome, and 
\cref{sect:irs} compares spectra simulated from it to actual \textit{Spitzer} spectra.
\cref{Sect:Discn} discusses our parameter constraints and the implications of the results.
\cref{Sect:Conclusion} summarizes our findings.
  \section{Observations} \label{Sect:Obs}
\begin{figure}  
    \centering
    \includegraphics[width=\columnwidth]{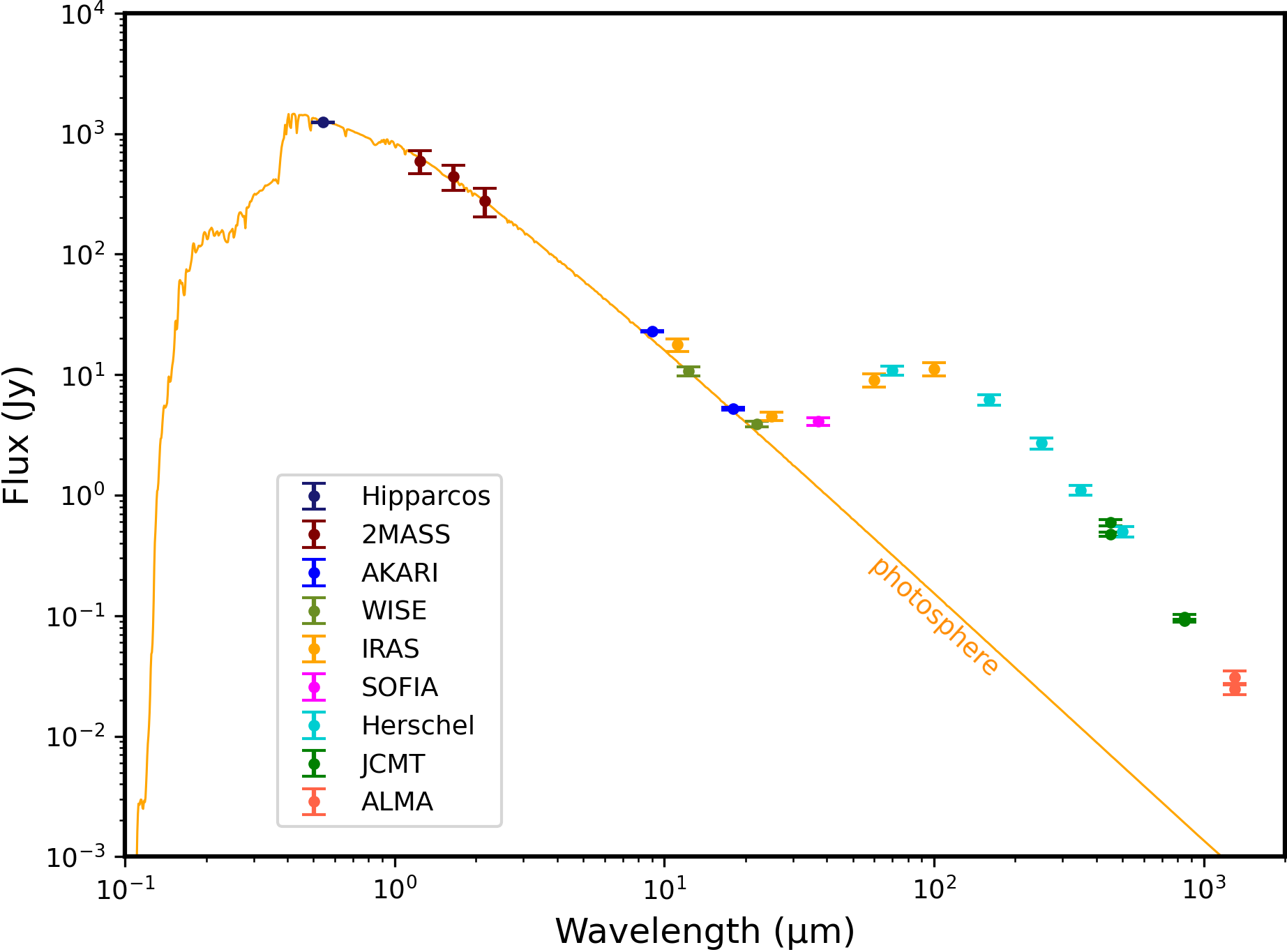}
    \caption{Fomalhaut photometry, including our fitting observables in the mid-infrared to sub-mm 
    (\cref{tab:phot_data}), as well as datapoints at smaller wavelengths used in the fitting
    of our photosphere model.}
    \label{fig:SED_phoenix}
\end{figure}

In this section, we examine the observational data utilized to fit our analytical model
of the Fomalhaut debris disk. 
The data include overall excess flux measurements at different wavelengths,
representing the disk's SED, 
which are summarized in \cref{tab:phot_data} and \cref{fig:SED_phoenix},
and the radial brightness distribution captured by JWST's Mid-Infrared Instrument (MIRI). 
The disk's SED encodes comprehensive information on the disk's overall thermal emission,
predominantly reflecting the characteristics of the larger,
colder particles in the outer belt, which emit in the far-infrared and (sub-) millimetre regime
and are less prone to PR drag migration.
Conversely, radial brightness profiles in the mid-infrared provide detailed spatial information 
about the smaller and warmer dust particles, 
which have a distribution more significantly influenced by PR drag.
Together, these datasets encapsulate the coupling of particle sizes and optical properties 
across the entire debris disk, which we aim to reproduce simultaneously.
ALMA observations are additionally used to inform our fixed model parameters concerning
the location of the dust-producing planetesimals, as discussed in \cref{Sect:ApplFom}.

\begin{table}
    \centering
    \caption{Photometric observations and excess emission from Fomalhaut.
        The excess fluxes ($F_\mathrm{disk})$ were obtained by subtracting
        the star's photosphere flux from total observed fluxes
        (except for ALMA-imaging-derived disk fluxes, denoted by \textdagger), 
        which are obtained from the corresponding references.
        For the photosphere fluxes, we use a stellar SED modelled with \texttt{PHOENIX} 
        \citep{Allard2012models} (provided by G.~Kennedy, pers.\@ comm., 2024),
        which was fitted using the method described in \citet{Yelverton2019statistically,Yelverton2020no}.}
        \label{tab:phot_data}
    \begin{threeparttable}
    \begin{tabular}{c c c c c}
        \hline\\[-1em]
        $\lambda$ (\si{\um}) & $F_\mathrm{total}$ (\si{\jansky}) & $F_\mathrm{disk}$ (\si{\jansky}) & Observatory \\[0.1em]
        \hline\hline\\[-1em]
        18   & \num{5.338\pm.114}  & \num{0.348\pm0.117}  & AKARI\tnote{1}    \\ 
        23   & \num{4.04\pm.22}    & \num{0.596\pm0.213}  & WISE\tnote{2}          \\ 
        25   & \num{4.81\pm.385}   & \num{0.867\pm0.385}  &  IRAS\tnote{3}   \\ 
        37.1 & \num{4.10\pm.28}    & \num{2.94\pm0.28}    & SOFIA\tnote{4}     \\ 
        60   & \num{9.02\pm1.08}   & \num{8.433\pm1.08}   & IRAS\tnote{3}   \\ 
        70   & \num{10.8\pm.90}    & \num{10.47\pm.90}    & Herschel\tnote{5}    \\ 
        100  & \num{11.2\pm1.46}   & \num{11.03\pm1.46}   & IRAS\tnote{3}  \\ 
        160  & \num{6.20\pm.6}     & \num{6.14\pm0.6}     & Herschel\tnote{5} \\ 
        250  & \num{2.70\pm.3}     & \num{2.68\pm0.3}     & Herschel\tnote{5} \\ 
        350  & \num{1.10\pm.1}     & \num{1.088\pm0.1}    & Herschel\tnote{5} \\ 
        450  & \num{.595\pm.035}   & \num{.588\pm.035}    & JCMT/SCUBA\tnote{6}     \\ 
        450  & \num{.475\pm.021}   & \num{.468\pm.021}    & JCMT/SCUBA-2\tnote{7}    \\ 
        500  & \num{.500\pm.050}   & \num{.494\pm.050}    & Herschel\tnote{5}    \\ 
        850  & \num{.097\pm.005}   & \num{0.095\pm0.005}  & JCMT/SCUBA\tnote{6}    \\ 
        850  & \num{.0912\pm.0025} & \num{0.089\pm0.0025} & JCMT/SCUBA-2\tnote{7}    \\ 
        1300 & -- & \num{0.0247\pm0.0025}\tnote{\textdagger} &  ALMA\tnote{8}  \\ 
        1300 & -- & \num{0.0308\pm0.0041}\tnote{\textdagger}$\;\,$\tnote{\textasteriskcentered} & ALMA\tnote{9} \\ 
        \hline
    \end{tabular}
    \begin{tablenotes}
        \item[\phantom{0}] 
        $^1$ \citet{Ishihara2010akari}, 
        $^2$ \citet{Wright2010wide},
        $^3$ \citet{Helou1988infrared},
        $^4$ \citet{Adams2018dust},
        $^5$ \citet{Acke2012herschel},
        $^6$ \citet{Holland2003submm},
        $^7$ \citet{Holland2017sons},
        $^{8}$ \citet{MacGregor2017complete},
        $^{9}$ \citet{White20173mm}
        \item[\textdagger] Direct fluxes from just the outer planetesimal belt.
        \item[\textasteriskcentered] 95\% confidence interval instead of $1\sigma$ error.
    \end{tablenotes}
    \end{threeparttable}
\end{table}

{
\setlength{\tabcolsep}{.1pt}
\begin{figure*}
    \centering
    \begin{tabular}{cccc}
    F1550C & F2300C & F2550W\\
    \includegraphics[width=.3\textwidth,trim={0 0 0 0},clip]{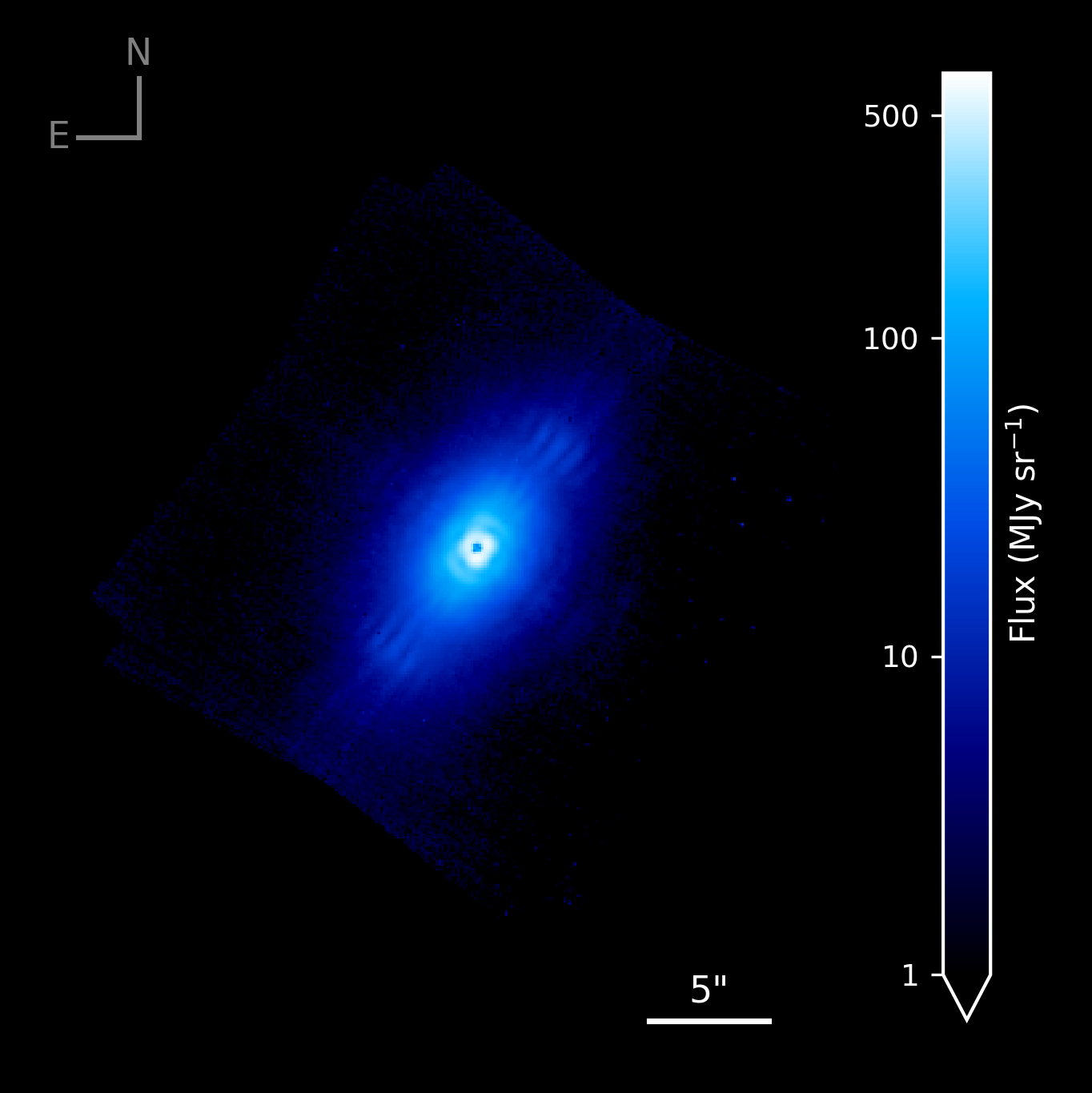}&
    \includegraphics[width=.3\textwidth,trim={0 0 0 0},clip]{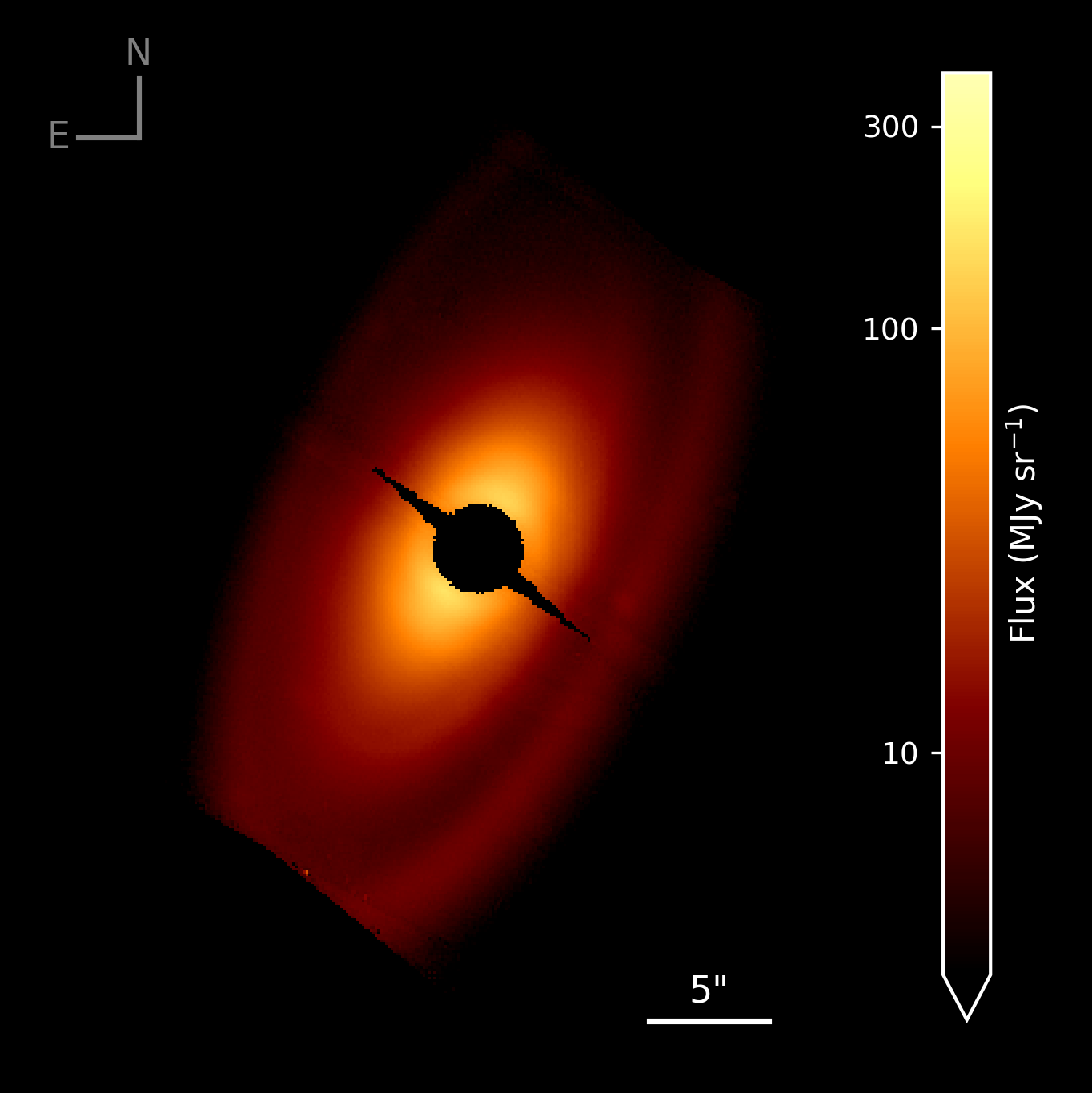}&
    \includegraphics[width=.3\textwidth,trim={0 0 0 0},clip]{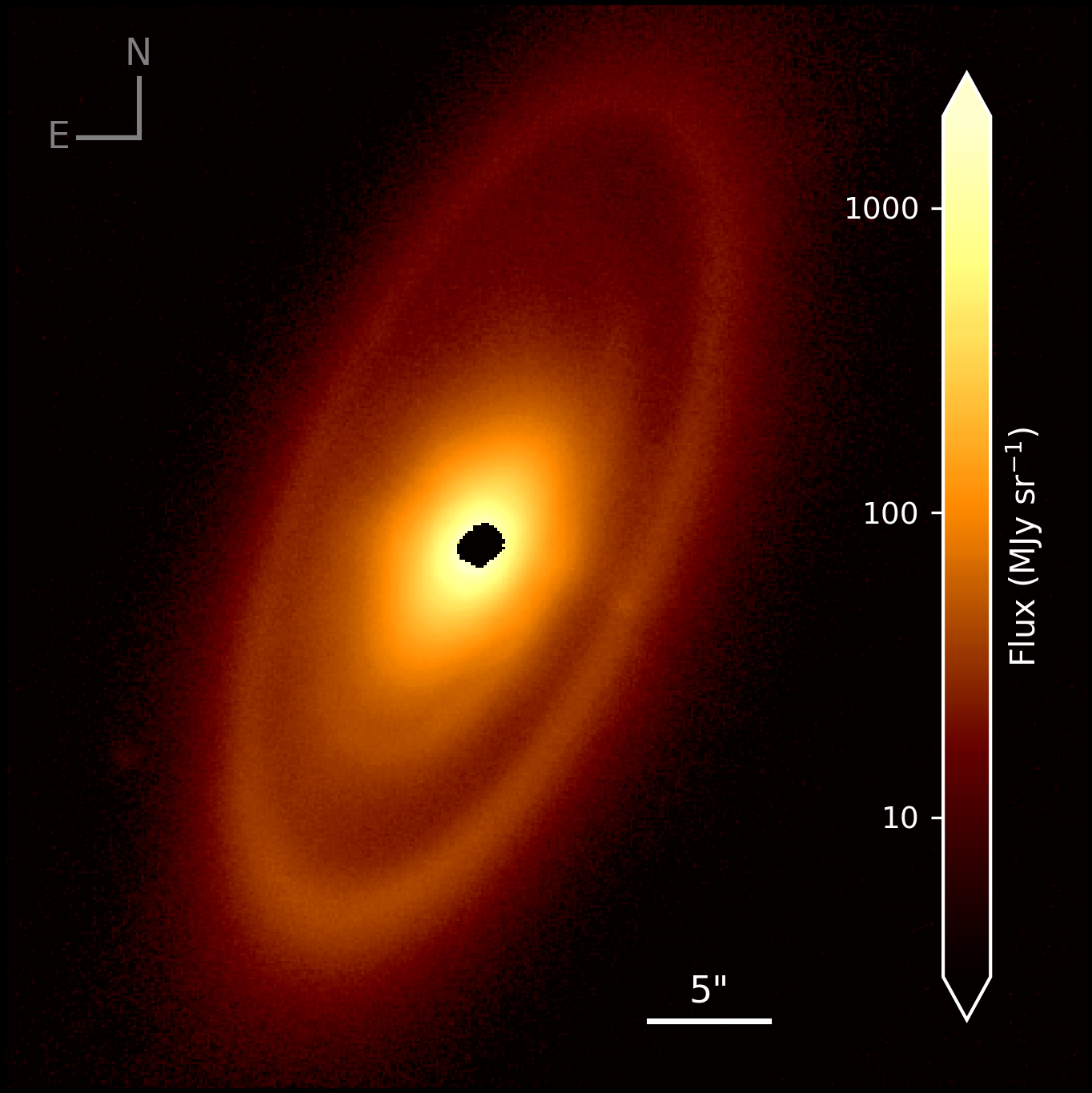}
    \end{tabular}
    \caption{JWST/MIRI images of the Fomalhaut system reproduced from the publicly available data from 
    \citet{Gaspar2023spatially},
    with their additional scaling factor of \num{3.14} for the F2300C image reversed
    (discussed in \cref{sect:best_fit}).
    Total image fluxes are \qty{89\pm57}{\milli\jansky}, \qty{176\pm50}{\milli\jansky}, and 
    \qty{895\pm73}{\milli\jansky} (from left to right; uncertainties extracted from 
    \citeauthor{Gaspar2023spatially}, \citeyear{Gaspar2023spatially}, their Fig.~3).}
    \label{fig:MIRI_images}
\end{figure*}
}

We extract the mid-infrared radial surface brightness profiles from the high-resolution images
of the Fomalhaut disk recently obtained by JWST/MIRI at three different wavelengths
with the \qty{15.5}{\um} and \qty{23}{\um} coronagraphic filters (F1550C \& F2300C)
as well as the \qty{25.5}{\um} filter (F2550W) \citep{Gaspar2023spatially}.
These images are reproduced in \cref{fig:MIRI_images} from the published data.
Following \citet{Gaspar2023spatially}, we deproject the reduced images using 
a position angle (PA) of \qty{336.28}{\degree} and an inclination of \qty{67.52}{\degree}, 
and then find the median surface brightness at bins of astrocentric distance.
Due to the unavailability of detailed error data, we use the standard deviation
of the azimuthal variation at each given radius as a proxy for the uncertainty.
Given an apparent inconsistency of the F2300C reduced fluxes with previously taken spectra,
\citet{Gaspar2023spatially} suspect an issue with the F2300C calibration pipeline
(see their Supplementary Section 1.2).
Therefore, while we display the F2300C data for comparison with our models,
we do not use this data for model fitting.

  \section{Disk Modelling} \label{Sect:Method}
\subsection{PR drag disk model} \label{Sect:model}
We employ the analytical model developed by \citet{Rigley2020dust} 
to simulate the two-dimensional dust distribution---across radial extent and particle size---in
a disk formed by a collisionally grinding planetesimal belt.
The model combines previous analytical approaches to determine (1) the size distribution 
arising within the radial confines of the planetesimal belt under collisions and PR drag loss \citep{Wyatt2011debris},
as well as (2) the radially dependent size distribution interior to the belt 
under further collisional and drag-induced evolution \citep{Wyatt2005insignificance}.
The model reasonably approximates the results of numerical kinetic models 
\citep[\eg][]{Lieshout2014nearinfrared} whilst being significantly faster, 
thus enabling exploration of large parameter spaces.
For more details about the model the reader is referred to \citet{Rigley2020dust}.

We compute the dust grains' material-dependent optical properties, including their $\beta$-factors---the
ratio of radiation pressure to stellar gravity that governs a particle's PR drag evolution---and
thermal emission in a manner similar to \citet{Wyatt2002collisional}.
Following \citet{Li1997unified}, the method treats dust as aggregates of core-mantle grains,
applying Maxwell-Garnett effective medium theory to derive the optical constants 
for the composite material.
We then determine the grains' optical efficiency coefficients
using Mie theory, Rayleigh-Gans theory, or geometric optics,
in the respective wavelength regimes.
Finally, the absorption coefficients define the dust grain temperatures, which are computed by
balancing the absorbed stellar radiation with the emitted thermal radiation.

The PR drag disk model has previously been applied to target systems of the HOSTS survey 
to assess if sufficient quantities of warm inner dust---required to explain 
a number of exozodi detections by the LBTI nulling interferometer 
\citep{Ertel2020hosts}---can be sustained by material being dragged inwards from an outer belt,
or if alternative dust transport scenarios are necessary (\eg{} via comets).
In 5 of 9 systems with known outer belts, this `PR-drag-transport-only' model
could reasonably reproduce the detected exozodi levels \citep{Rigley2020dust}.
Similarly, here we apply this model to the Fomalhaut system,
while leveraging the more extensive observational data recently gathered by JWST/MIRI.

Both components of this model, that is, 
our implementations of the PR drag disk model from \citet{Rigley2020dust} and
of the method to find the dust optical properties and temperatures, 
are publicly available as individual \texttt{Python} packages 
(see Section \hyperref[Sect:DataAvailability]{Data Availability}).

\subsection{Synthesizing MIRI images}
For an adequate comparison of our model outcomes with the MIRI data, we need to generate
synthetic MIRI images from the simulated dust spatial distribution and optical properties.
Whereas \citet{Rigley2020dust} focussed on the total emission from the disk 
(which we also employ in form of the SED),
here we obtain the spatially resolved emission by first determining the radial profile 
of the disk's face-on surface brightness at a specific wavelength as
\begin{align} \label{eq:Fnu}
    S_\nu (r) = \int Q_\mathrm{abs}(\lambda, D) \; B_\nu[T(D,r)] \; \tau(D,r) \, \mathrm{d}D \;\;,
\end{align}
where $Q_\mathrm{abs}(\lambda, D)$ are the grains' absorption efficiencies, 
$B_\nu$ their spectral radiance, $T(D,r)$ their temperatures, 
and $\tau(D,r) \mathrm{d}D$ is the geometrical depth in particles of size 
$D\!\rightarrow\!D\!+\!\mathrm{d}D$ at radius $r$.

For a given face-on radial brightness profile, disk inclination, and scale height aspect ratio,
we produce the corresponding astrophysical scene
(\ie{} how the inclined disk would appear to an ideal observer, free from observational distortions) 
using the \rave{} package \citep{Han2022rave}.
This method draws sample points that are vertically Gaussian, 
azimuthally uniform, and radially distributed according to the radial profile, 
before binning them into a pixel grid projected at the appropriate orientation in space.
We adopt a pixel scale of \qty{0.26}{\arcsecond} per pixel
(corresponding to \qty{2}{\astronomicalunit} at the distance of Fomalhaut),
slightly more than twice the native scale of MIRI (\qty{0.11}{\arcsecond} per pixel).
This coarser sampling serves to limit the computational costs
imposed by the point spread function (PSF) convolution step for the coronagraphic filters,
where the PSF varies across the field of view.

To produce synthetic MIRI images, we convolve the astrophysical scenes 
with the PSF of the respective MIRI filter, obtained using the \webbpsf{} package 
\citep[v.~1.3.0; renamed \texttt{STPSF} as of v.~2.0,][]{Perrin2012simulating}.
For the coronagraphic filters, where the PSF varies across the image, 
we pre-generate a data cube storing the PSF for each individual pixel,
requiring \qty{1.3}{\giga\byte} and \qty{2.0}{\giga\byte}
of memory for F1550C and F2300C, respectively.

We consider only the emission at the central wavelength of each filter.
Extending our pipeline to multiple wavelengths per filter, 
which would also require caching the spatially variant PSFs for each wavelength, 
would be computationally prohibitive for our purpose.
Given the bandwidth of the filters, this simplification might introduce inaccuracies.
However, we found that, at least for the unconvolved face-on radial brightness profiles,
the difference between central-wavelength and band-integrated profiles
\citep[with wavelength-dependent throughputs for the latter obtained from \textit{Pandeia}, 
the JWST exposure time calculator system;][]{Pontoppidan2016pandeia}
is negligible for our purpose, reaching at most a few percent.
To replicate the saturation at the stellar core in the observations, 
we mask the pixels within \qty{1.2}{\arcsecond} in the F2550W image.
No noise is added to the images.
We also do not replicate the differential imaging in the observations.
Only one synthetic image is generated for each filter 
with the detector's vertical axis aligned with the disk's minor axis,
comparable with the observations.

Since we model only the axisymmetric radial distribution of dust,
we do not compare the observed and synthetic images directly 
but instead the radial brightness profiles extracted from the deprojected images,
thereby ignoring any azimuthal features observed in the actual disk, 
in particular the slight eccentricity of the outer planetesimal belt, as well as the eccentric 
`intermediate ring' feature superimposed on the disk \citep{Gaspar2023spatially}.
For deprojection, we again use the disk PA and inclination found by \citet{Gaspar2023spatially},
so that we are comparing profiles extracted from the observations and simulated model images
in exactly the same way.

\subsection{Application to Fomalhaut} \label{Sect:ApplFom}
To apply the PR drag model to the Fomalhaut disk, it is necessary to set appropriate input 
parameters for the central star, the planetesimal belt, and the particle characteristics.

For the central star, we adopt a mass $M_{\star}\!=\!\qty{1.92}{\solarmass}$ \citep{Mamajek2012age},
and a luminosity $L_{\star}\!=\qty{16.6}{\solarlum}$ and spectrum (shown in \cref{fig:SED_phoenix}, 
$T_\mathrm{eff}\!=\!\qty{8523}{\kelvin}$) modelled with \texttt{PHOENIX} \citep{Allard2012models},
as found via the SED fitting method by \citet{Yelverton2019statistically,Yelverton2020no}
(spectrum provided by G.~Kennedy, pers.\@ comm., 2024).

For the planetesimal belt, the model requires inputs for its geometric characteristics---namely, 
its radial extent and opening angle---as well as its dust mass, 
which is defined as the total mass up to a particle size of \qty{1}{\cm},
the largest size included in the model.
The radial dimensions of the belt are informed by millimetre imaging data from ALMA observations
\citep{MacGregor2017complete}, which delineate the location of the planetesimals. 
These data suggest a moderately eccentric belt ($e\!=\!\num{0.12\pm0.01}$) with
an inner edge at pericentre at \qty{119.9 \pm 0.8}{\astronomicalunit} and
an outer edge at apocentre at \qty{152.6 \pm 1.0}{\astronomicalunit}.
Given that our model is axisymmetric, we approximate the observed belt as a wider, 
circular ring with an inner and outer edge fixed at \qty{119.9}{\astronomicalunit} 
and \qty{152.6}{\astronomicalunit}, respectively, 
thereby preserving the radial extent of the actual belt.

The opening angle of the disk may also be estimated from the ALMA imaging data.
We employ a parametric belt model to fit the full ALMA map by \citet{MacGregor2017complete},
ignoring the belt's eccentricity and instead assuming an axisymmetric ring 
centred on the geometric centre of the projected disk emission. 
Further assuming a radially and vertically Gaussian ring with peak intensity, peak radius, 
radial width, inclination and scale height aspect ratio as free parameters
we find a vertical half-opening angle of \qty{1.4 \pm .1}{\degree} under this simplified model. 
This is comparable to the half-opening angle of \qty{1.0 \pm .25}{\degree} found by \citet{Boley2012constraining}
by fitting an axisymmetric ring to a partial ALMA map the disk, 
although a vertically exponential profile was assumed in their model
rather than a Gaussian profile.
On the other hand, \citet{Kennedy2020unexpected} find a half-opening angle of \qty{1.7}{\degree} 
derived from fitting particle orbit distributions to the full ALMA map.\footnote{
    A coding error led to an overly loose constraint on particle inclinations 
    in \citet{Kennedy2020unexpected}. 
    The corrected value for the half-opening angle given here was provided 
    by G.~Kennedy (pers.\@ comm., 2024).
}
For simplicity, here we adopt a value of \qty{1.5}{\degree}.
We discuss how variations of this parameter affect our model outcomes in \cref{Sect:collstr}.

The original model assumed the collisional velocity to be dominated by 
the particles' vertical motion, with $v_\mathrm{rel}\!=\!v_\mathrm{k} I_\mathrm{max}$ 
\citep[][Eq.~5]{Rigley2020dust}, where $v_\mathrm{k}$ is the circular Keplerian velocity 
and $I_\mathrm{max}$ is the particles' maximum inclination in radian, 
corresponding to the semi-opening angle of the disk.
To also account for an eccentricity distribution within the eccentric belt,
we add a velocity component depending on the particles' mean proper eccentricity;
$v_\mathrm{rel}\!=\!v_\mathrm{k} \sqrt{1.25 \, \smash{e_\mathrm{p}^{\scriptscriptstyle 2} 
\,+\, I_\mathrm{max}^{\scriptscriptstyle 2}}}$ \citep{Lissauer1993growth}.
For the mean proper eccentricity, $e_\mathrm{p}$, we adopt a value of 0.019, as found by 
\citet{Kennedy2020unexpected}.

The belt dust mass---unlike other parameters that may be varied freely or fixed from the start---is
set individually for each run such that the resulting total flux from the belt at 
$\lambda\!=\!\qty{1.3}{\mm}$ matches the belt flux measured by ALMA at that wavelength.
Specifically, we scale the mass to match the mean of the fluxes found by
\citet{MacGregor2017complete} via fitting in the image plane (\qty{24.7 \pm 0.1}{\milli\jansky}), and by
\citet{White20173mm} via fitting the visibility data (\qty{30.8 \pm 0.1}{\milli\jansky}).
The mass is found by rerunning the model iteratively, until the difference
between the modelled and measured flux is \qty{<1}{\percent} 
(usually achieved within three iterations).

We assume the PR-drag-maintained disk to extend down to \qty{0.5}{\astronomicalunit}.
This is roughly comparable to the sublimation distance of refractory dust
\citep[\qtyrange{0.2}{0.3}{\astronomicalunit} for silicate,][]{Lebreton2013interferometric}.
The observables we simulate and compare in this study---mid-infrared up to sub-mm wavelengths---proved
relatively insensitive to the exact location of the innermost edge of the disk,
provided it is $\lesssim\!\qty{1}{\astronomicalunit}$.

\subsection{Particle parameters}

The particle parameters in our model pertain to the grain composition and structure, 
which determine their optical properties, as well as to the grains' collisional strength,
and are treated as free parameters.
The dust grain model we employ \citep{Li1997unified} assumes that particles 
are porous aggregates of core-mantle grains with a silicate core 
and an organic refractory mantle, as well as water ice filling the voids.
It is parametrized by three fractions:
(1) The volume fraction of silicate in the combined volume of silicates 
and organic refractory material, $\qsi = v_\mathrm{Si} / ( v_\mathrm{Si} + v_\mathrm{C} )$.
(2) The matrix porosity, \por{}, defined as the volume fraction of voids
within the total volume of the silicate/organic matrix. These voids may be empty
or partially filled with water ice, as determined by (3) the \qwa{} parameter,
defined as the ratio of ice volume to total void volume.
For each of these parameters, we test the full range (\ie{} \numrange{0}{1}, 
or rather \numrange{0}{.99} in the case of porosity).

For the collisional strength, the model of \citet{Rigley2020dust} adopts a
power-law determining the critical specific energy \qdstar{}, that is,
the kinetic energy threshold required for an impactor to catastrophically disrupt a target,
following \citet{Benz1999catastrophic}.
This two-parameter prescription is of the form $\qdstar = \qdnorm ( D / \si{\cm} ) ^ {\qdslope}$,
with a normalization parameter, \qdnorm{}, and a slope parameter, \qdslope{}.
Conventionally assumed parameters for \qdstar{}
typically correspond to a critical specific energy for a \qty{1}{\cm} target (\ie{} \qdnorm)
within one order of magnitude from \qty{e7}{\erg\per\g}
(\eg{} \cite{Krivov2006dust}: \qty{3.02e6}{\erg\per\g},
\cite{Grun1985collisional}: \qty{\sim9e6}{\erg\per\g},\footnote{
    As retrieved by \citet{Pokorny2024how} from the alternative expression
    for \qdstar{} by \citet{Grun1985collisional}.}
\cite{Heng2010longlived}: \qty{e7}{\erg\per\g},
\cite{Lohne2008longterm}: \qty{2.45e7}{\erg\per\g}).
However, a recent study of the collisional lifetimes of solar system meteoroids
finds $\sim\!\!3$ orders of magnitude higher collisional strengths
than conventionally assumed are necessary to fit observed meteor dynamics \citep{Pokorny2024how}.
To incorporate this large span of estimated strengths, 
we choose a dynamic range of a factor of \num{e4} for \qdnorm{}.
For \qdslope{}, we explore a range from $-0.9$ to $0.45$,
comfortably accommodating the span of typically assumed values of around \numrange[range-phrase={ to }]{-0.3}{0}
\citep[\eg][]{Krivov2006dust,Lohne2008longterm,Heng2010longlived}.
We distinguish two fitting scenarios; 
Scenario~A, across the entire range of tested values for \qdslope{}, and
Scenario~B, where only the more conventional range of $\qdslope\!\leq\!0$ is considered.

With 9--14 values per parameter tested (see \cref{tab:parameter}),
this amounts to \num{131670} parameter combinations, 
including \num{1463} unique compositions
after excluding redundant settings (\ie{} variations of \qwa{} for $\por\! =\!0$).
We precompute the optical properties and radially dependent temperatures for the \num{1463} materials
across 45 particle sizes logarithmically spaced between \qty{4}{\um} and \qty{1}{\cm},
taking around 1~hour on a server with a 40-core processor.
We then run the PR drag model, including the generation of the three synthetic MIRI images,
with the full parameter grid, taking around 6~hours on the same machine.
The per-core computing time of each individual model is around 6 seconds,
with most of this time dedicated to image generation and convolution.

{
\addtolength{\tabcolsep}{-1pt}
\begin{table}
    \centering
    \caption{Parameter space for fitting.} \label{tab:parameter}
    \begin{tabular}{ m{1.5cm} c c l }
    \hline\\[-1em]
    Parameter           & Range                             & Count & Scale   \\[0.1em]
    \hline\hline\\[-1em]
    \qsi        & \numrange{0}{1}                           & 11    & linear            \\
    \qwa        & \numrange{0}{1}                           & 11    & linear            \\
    \por        & \numrange{0}{.9}, \numrange{.95}{.99}     & 10, 3 & lin. (2 regimes)  \\
    \qdnorm   $\;\;\left[\unit{\erg\per\g}\right]$   
                & \numrange{e6}{e9}                         & 9     & log.              \\
    \qdslope    & \numrange{-0.9}{0.45}\hspace{.6mm}\text{} & 10    & linear           \\
    \hline
\end{tabular}
\end{table}
}

  \section{Model Fitting} \label{Sect:Fitting}
\subsection{Goodness-of-fit metric} \label{Sect:GOF}
To quantify how well the model reproduces the observations under a given set of
parameters, we construct goodness-of-fit functions $\chi^2$,
where the best fit has the minimum $\chi^2$.
To simultaneously match both our observational datasets reasonably well, 
we will introduce $\chitot$ as a combination of $\chised$ and $\chirad$,
which are the respective goodness-of-fit metrics
for the SED and for the MIRI-retrieved radial brightness profiles.

We start from the general form of the $\chi^2$ statistic used to quantify the goodness-of-fit 
between observed data and model predictions, which is given by:
\begin{align}
\chi^2 = \sum_{i} \left( \frac{O_i - E_i}{\sigma_i} \right)^2 \;,
\end{align}
where $O_i$ are the observed values, $E_i$ are the expected (model) values, 
and $\sigma_i$ are the uncertainties in the observed values. 
To appropriately handle the logarithmic nature of our data
(both in values, \ie{} flux, and in sampling points, 
\ie{} wavelength or, respectively, radial distance from the star), 
we first transform the observed and model values, 
as well as the uncertainties, to logarithmic scales, such that we replace:
\begin{align}
O_i \rightarrow \log_{10}(O_i), \quad E_i \rightarrow \log_{10}(E_i), \quad \sigma_i \rightarrow \frac{\sigma_i}{O_i \ln(10)} \;.
\end{align}

In addition, given that the observational data points may not be evenly spaced, 
we incorporate a weighting mechanism based on data point density in logarithmic space.
We calculate weights as the inverse of the distance between preceding and following data points:
\begin{align}
w_i = \frac{1}{\Delta x_{\log, i}}, \quad \text{where} \quad \Delta x_{\log, i} = \log_{10}(x_{i+1}) - \log_{10}(x_{i-1}) \;,
\end{align}
where $x_i$ are the sampling points of the observed values.
This ensures that closely-spaced (in log-scale) data points do not excessively dominate the fit.
We obtain:
\begin{align}
\chi^2 = \sum_{i} w_i \left( \frac{\log_{10}(O_i) - \log_{10}(E_i)}{\sigma_{\log_{10}(O_i)}} \right)^2 \;,
\end{align}
where $\sigma_{\log_{10}(O_i)} = \sigma_i / (O_i \ln(10))$.

In the case of the two radial profiles retrieved with the F1550C and F2550W filters
(note that we omit the F2300C profile, see \cref{Sect:Obs}),
we compute one $\chi^2$ value for each and add them up to form \chirad{}:
\begin{align}
\chi^2_{\text{rad}} = \chi^2_{\text{rad, F1550C}} + \chi^2_{\text{rad, F2550W}}  \;.
\end{align}

Lastly, to ensure that both datasets equitably contribute to the combined goodness-of-fit metric,
we normalize the `unscaled' $\chi^2$ metrics by dividing them by their respective best-fit values
across all parameter sets:
\begin{align}
\chi^2_{\text{norm, SED}} = \frac{\chised}{\chi^2_{\text{best, SED}}}, \quad \chi^2_{\text{norm, rad}} = \frac{\chirad}{\chi^2_{\text{best, rad}}} \;.
\end{align}

This normalization is necessary for two reasons.
Firstly to factor in the varying numbers of sampling points in both observational datasets---compounded
by the fact that multiple radial profiles were retrieved---and, 
secondly, to account for the different magnitudes of the uncertainties accompanying the datasets.
The stated uncertainties for the SED are roughly an order of magnitude
smaller than the uncertainties of the radial profiles,
for which we use the azimuthal variation at a given radius (see \cref{Sect:Obs}).
This normalization essentially equalizes the uncertainty scales of the two datasets,
while upholding the relative weighting effect the uncertainties have
among data points within each dataset.

Finally, we define the combined goodness-of-fit metric as the mean 
of the normalized $\chi^2$ values for each dataset:
\begin{align}
\chitot = \frac{ \chi^2_{\text{norm, SED}} + \chi^2_{\text{norm, rad}} }{2} \;.
\end{align}
This metric approaches one for the best achievable outcome, although 
it can only reach one if the best fits for both datasets are achieved simultaneously. 
Due to the normalization, this metric does not signify absolute goodness-of-fit
but rather represents a relative measure of model performance that allows us to
easily compare different model configurations across the entire parameter grid.

\begin{figure*}
    \centering
    \includegraphics[width=.8\textwidth,trim={0 0 0 0},clip]{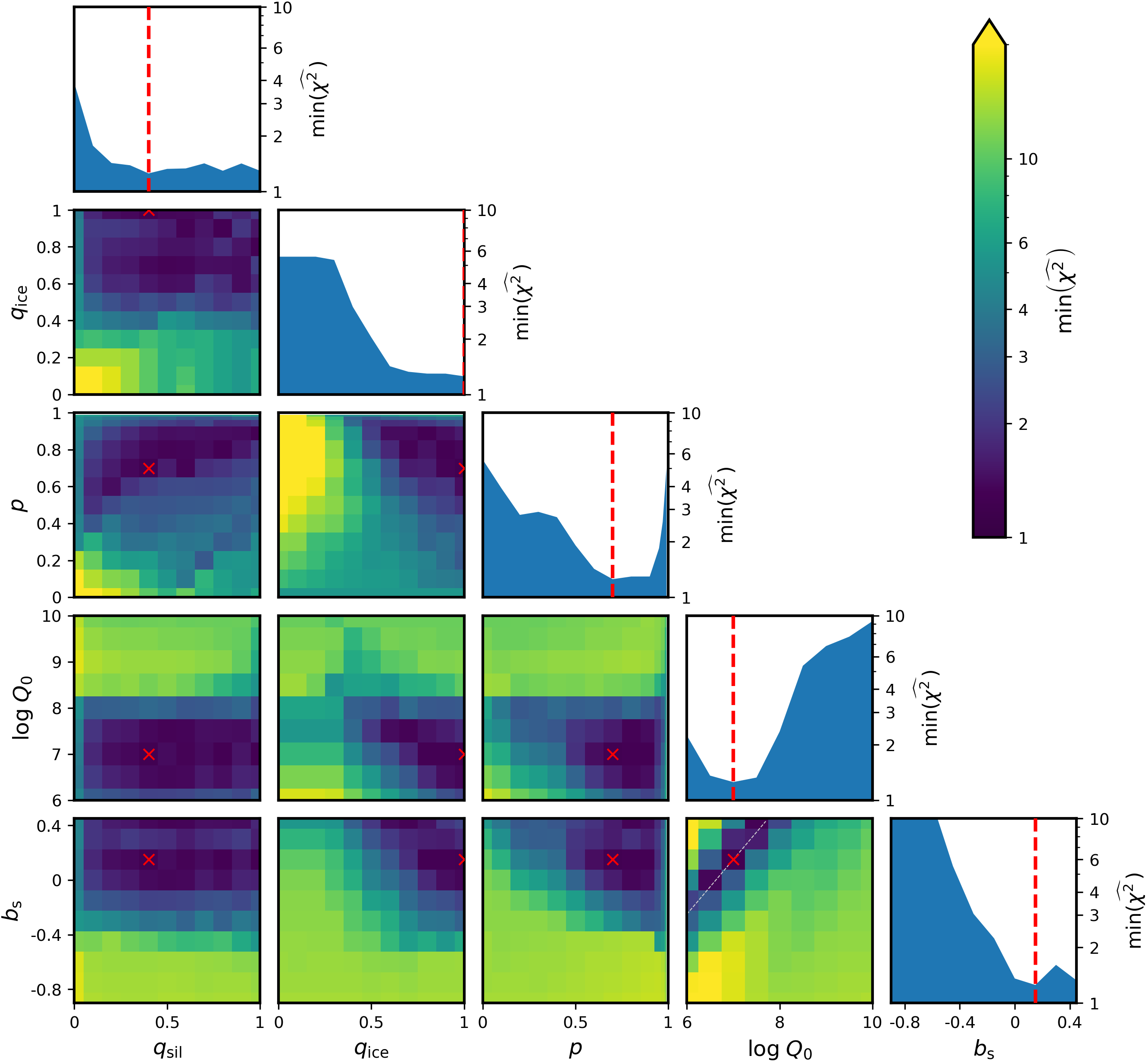}
    \caption{Variation of goodness-of-fit metric, \chitot{}, across parameter space.
    Contour plots (lower triangle) display the minimum achieved \chitot{}
    for any two-parameter combination.
    Plots on the diagonal show the minimum achieved \chitot{} across the range of each individual parameter.
    Red crosses and lines mark the best fit (Scn.~A).}
    \label{fig:corner}
\end{figure*}

\subsection{Exploring the parameter space for good fits}
To explore which combinations of parameters optimize the fit
between our model predictions and the observational data, 
and to identify interactions among these parameters,
we assess the \chitot{} distribution across our models using a corner plot (\cref{fig:corner}). 
The pseudocolour diagrams in the lower triangle of the plot illustrate the interplay
of any two parameters by showing the minimum achieved \chitot{}
for each pairing while marginalizing over the remaining parameters.
This approach helps us identify regions within the parameter space where the model fit is most favourable. 
The plots on the diagonal highlight how \chitot{} varies
as each parameter is adjusted independently, 
indicating the sensitivity of the model fit to changes in individual parameters.
By utilizing a \chitot{} cutoff, we may establish confidence intervals for these parameters.
Given the lack of an absolute goodness-of-fit measure, we empirically set a cutoff at $\chitot <\!1.5$, 
which we found to yield acceptable fits.
\cref{tab:best_fit} presents the best-fit parameters (those achieving the overall minimum \chitot{} value)
along with the corresponding uncertainties, 
defined by the range of parameter values that fall within our \chitot{} cutoff.

First looking into the dependence of the goodness-of-fit on individual parameters,
we note that suitable parameters suggest fairly icy particles---characterized
by high porosity and a high water ice filling factor---with
a weakly constrained silicate-organics ratio.
The conventional range for \qdnorm{} fits well,
while the higher collisional strengths found for solar system grains by \citet{Pokorny2024how} are rejected.
The commonly assumed slightly negative values for \qdslope{} do allow for a suitable fit,
yet the tendency for \qdslope{} to favour positive values is noteworthy.
To assess how a more conventional \qdstar{} power law slope would change our parameter constraints,
we can also explore Scenario B where we impose $\qdslope{} \leq 0$.
This constraint worsens the fit only marginally, yet tightens the requirements
for a high \qwa{} and low \qdnorm{} (see \cref{tab:best_fit}).

{
\addtolength{\tabcolsep}{-1pt}
\begin{table}
    \centering
    \caption{Parameters of suitable models.
    Scenario A includes all models,
    whereas Scenario B includes only those with a flat or negative \qdstar{} power-law slope, 
    \ie{} $\qdslope\!\leq\!0$.
    The upper five parameters are the free model parameters, 
    which directly determine the centre four `secondary' parameters.
    The bottom three parameters are quantities derived from the model outcomes.} \label{tab:best_fit}
    \begin{tabular}{ m{22mm} c c c c c }
    \hline\\[-1em]
    Parameter       & \multicolumn{2}{c}{Best fit}        && \multicolumn{2}{c}{Within \chitot{} cutoff}  \\[0.1em]
    \cline{2-3} \cline{5-6}\\[-1em]
    \hfill Scn.:    & A (all)   & B ($\qdslope\!\leq\!0$) && A (all)              & B ($\qdslope\!\leq\!0$)\\[0.1em]
    \hline\hline\\[-1em]
    \qsi            & \num{.4}  & \num{.5}                && \numrange{.2}{1.0}     & \numrange{.4}{.8}    \\
    \qwa            & \num{1.0} & \num{1.0}               && \numrange{.6}{1.0}     & \numrange{.9}{1.0}    \\
    \por            & \num{.7}  & \num{.8}                && \numrange{.6}{0.9}    & \numrange{.8}{.9}     \\
    $\log{\qdnorm}$ & \num{7}   & \num{6.5}               && \numrange{6.5}{7.5}    & \num{6.5}             \\
    \qdslope        & \num{.15} & \num{0}                 && \numrange{0}{0.45}     & \num{0}               \\[0.1em]
    \hline\\[-1em]
    $\qwatot = v_\mathrm{ice} \,/\, v_\mathrm{tot}$
                    & \num{.7}  & \num{.8}                && \numrange{.54}{.81}     & \numrange{.8}{.81}     \\
    $\portot = v_\mathrm{vac} \,/\, v_\mathrm{tot}$
                    & \num{.0}  & \num{.0}                && \numrange{.0}{0.36}    & \numrange{.0}{.09}    \\
    $(m_\mathrm{sil} + m_\mathrm{carb}) / m_\mathrm{ice}$
                    & \num{1.2} & \num{0.72}              && \numrange{.43}{1.6}    & \numrange{.43}{.77}    \\
    $\log{\qdstarcon}$
                    & \num{6.6} & \num{6.5}               && \numrange{6.4}{6.6}    & \num{6.5}             \\[0.3em]
    \hline\\[-1em]
    $\chitot$       & \num{1.25}& \num{1.36}              && \num{<1.5}             &  \num{<1.5}           \\
    $\etacon$       & \num{29}  & \num{44}                && \numrange{26}{44}      &  \numrange{43}{44}   \\
    zodi level      & \num{68}  & \num{58}                && \numrange{55}{91}     &  \numrange{56}{64}
\end{tabular}
\end{table}
}

\subsubsection{Parameter correlations}
\begin{figure} 
    \centering
    \includegraphics[width=\columnwidth,trim={0 0 0 0},clip]{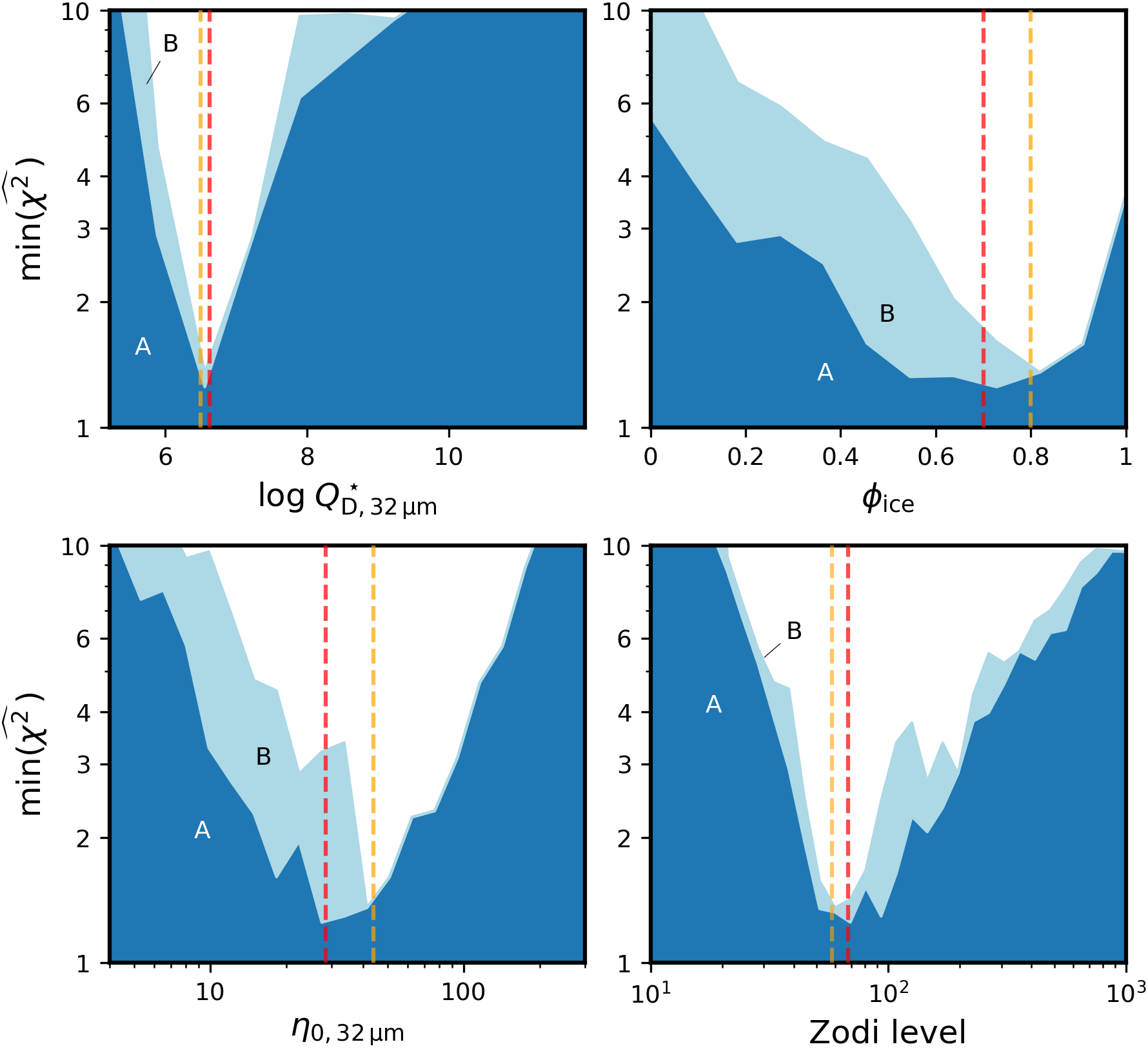}
    \caption{Goodness of fit across secondary parameters \qdstarcon{} and \qwatot{},
    as well as across the derived quantities \etacon{} and the zodi level.
    Shaded areas show the marginalized \chitot{} distribution including (Scn.~A) 
    and excluding (Scn.~B) models with positive \qdstar{} power-law slope.}
    \label{fig:derived_params}
\end{figure}

The marginalized \chitot{} distributions across two parameters,
seen in the lower triangle in \cref{fig:corner}, let us identify parameter correlations,
which manifest as skewed, elongated regions of suitable fits.
While most parameter combinations show no evidence for correlation, a pronounced correlation
between the collisional strength parameters \qdnorm{} and \qdslope{} is 
evident---apparently, $\qdslope\!\propto\!\log{\qdnorm}$.
We find that the region of suitable fits is described by a narrow corridor along a straight line
that is approximately described by:
\begin{align} \label{eq:qdline}
    \log{\qdnorm} = 2.5 \qdslope + 6.625
\end{align}
Along this line, \qdstar{} can thus be expressed as depending solely on \qdslope{}:
\begin{align}
    \qdstar = 10^{2.5 \qdslope + 6.625} \, \unit{\erg\per\g} \, (D/\unit{\cm})^{\qdslope}
\end{align}
Here, given the exponential growth of the terms $10^{2.5 \qdslope + 6.625}$ and $(D/\unit{\cm})^{\qdslope}$,
one may suspect that there exists a specific value of $D$, 
where the growth rates of these terms effectively counterbalance each other, 
resulting in a constant product \qdstar{} across different values of \qdslope{}.
To determine $D$ at which \qdstar{} is invariant with respect to variations in \qdslope{}, 
we differentiate $\log(\qdstar)$ with respect to \qdslope{} and set it to zero,
yielding $D\!=\!10^{-2.5}\,\unit{\cm}\!\approx\!\qty{32}{\um}$.
At this value, \qdstar{} is invariant for any combination of \qdnorm{} and \qdslope{}
that satisfies \cref{eq:qdline}, which is where the most suitable fits are found.
In other words, a favourable model outcome is less determined by \qdnorm{} or \qdslope{} per se,
but rather solely by \qdstarcon{},
which is optimal at around \qty{\sim4e6}{\erg\per\g}.
The marginalized distribution of \chitot{} over this derived parameter, \qdstarcon{},
is shown in \cref{fig:derived_params}~(top left). 
The similar distributions for Scenarios A and B
confirm that, for good fits, \qdstarcon{} is rather independent of \qdslope{}.
The fact that the size regime at which we obtain a constraint on \qdstar{}
is just above the blow-out size around Fomalhaut 
($\diambl\!:\!\beta(\diambl)\!=\!0.5,\; \diambl\!\approx\!\qty{13}{\um}$, 
for the best-fit material)
is consistent with the expectation that, in a collisional cascade, the geometrical depth
is dominated by the smallest particles that can remain bound, 
which therefore also contribute most to the emission we are fitting.

Another noteworthy parameter correlation exists between the porosity \por{}
and the ice filling fraction \qwa{}.
There appears to be a region, within which suitable fits may be found
that is roughly characterized by $\por\!\propto\!1/\qwa$.
This indicates a constraint on the product of \por{} and \qwa{},
which is the total ice volume fraction of the particle material,
$\qwatot = v_\mathrm{ice} \,/\, v_\mathrm{tot} = \por \cdot \qwa$.
Plotting the marginalized \chitot{} distribution over \qwatot{},
as given in \cref{fig:derived_params} (top right),
reveals an extended minimum spanning roughly $0.5\!<\!\qwatot\!<\!0.9$.
Considering only models with $\qdslope\!\leq\!0$ (Scn.~B), this constraint
tightens to $0.7<\!\qwatot\!<\!0.9$.

The constraints on these secondary model parameters, \qdstarcon{} and \qwatot, as well as the
overall grain porosity $\portot = v_\mathrm{vac} \,/\, v_\mathrm{tot} = \por \cdot (1 - \qwa)$,
are also summarized in \cref{tab:best_fit}. 

\subsubsection{Constraints on lifetimes and zodi level}
We may also analyze how certain derived quantities are constrained by the fitting process. 
The respective marginalized \chitot{} distributions are depicted in \cref{fig:derived_params} (bottom). 

For one, the constraint we obtained on \qdstarcon{} suggests that the fitting actually
constrains the collisional lifetimes of particles of that size---or more specifically,
the ratio of their PR drag lifetimes 
(the time it takes particles to migrate from the belt to the star)
to their collisional lifetimes (were they to remain in the belt), $\eta_0$.
This ratio, which depends on \qdstar{} among other factors, is a crucial intermediate quantity within the model
that influences the determination of the geometrical depth profile.
We find that the \etacon{} of good fits is on the order of several tens, 
with Scenario B favouring a slightly higher \etacon{} than Scenario A.

Another interesting parameter we can derive from the models is the zodi level.
This quantity is defined as the disc's surface density at the Earth-equivalent insolation distance 
(EEID)---calculated as $r_\mathrm{\,EEID} = \sqrt{L_\star/L_\odot}\,\si{\astronomicalunit}$,
approximately \qty{4.1}{\astronomicalunit} for Fomalhaut---relative to the surface density
of the solar system's zodiacal cloud at \qty{1}{\astronomicalunit} \citep{Kennedy2015exozodi}.\footnote{
    For the surface density of the zodiacal cloud at \qty{1}{\astronomicalunit} we use
    the value of \qty{7.12e-8}{\astronomicalunit^2/\astronomicalunit^2}, as retrieved
    by \citet{Kennedy2015exozodi} from the \citet{Kelsall1998cobe} model.
} 
As seen in \cref{fig:derived_params} (bottom-right), 
well-fitting models have a zodi level of around \num{60}, meaning that,
at the radius where a planet would receive the same energy as Earth does from the Sun,
the surface brightness is approximately \num{60} times 
that of the zodiacal cloud at the orbit of Earth. 
There is no significant difference between Scenarios A and B concerning this parameter.
However, we do not expect the disk at \qty{4}{\astronomicalunit} to strongly influence 
the observables we are fitting, making this prediction more of an extrapolative indicator
of disk density rather than a direct observational constraint.

  \section{The best-fit model} \label{sect:best_fit}
{
\setlength{\tabcolsep}{.1pt}
\begin{figure*}
    \centering
    \begin{tabular}{ccccc}
    & \phantom{a.} &F1550C & F2300C & F2550W\\
    \rotatebox{90}{\hspace{10mm} \Large before convolution}& &
    \includegraphics[width=.3\textwidth,trim={0 0 0 0},clip]{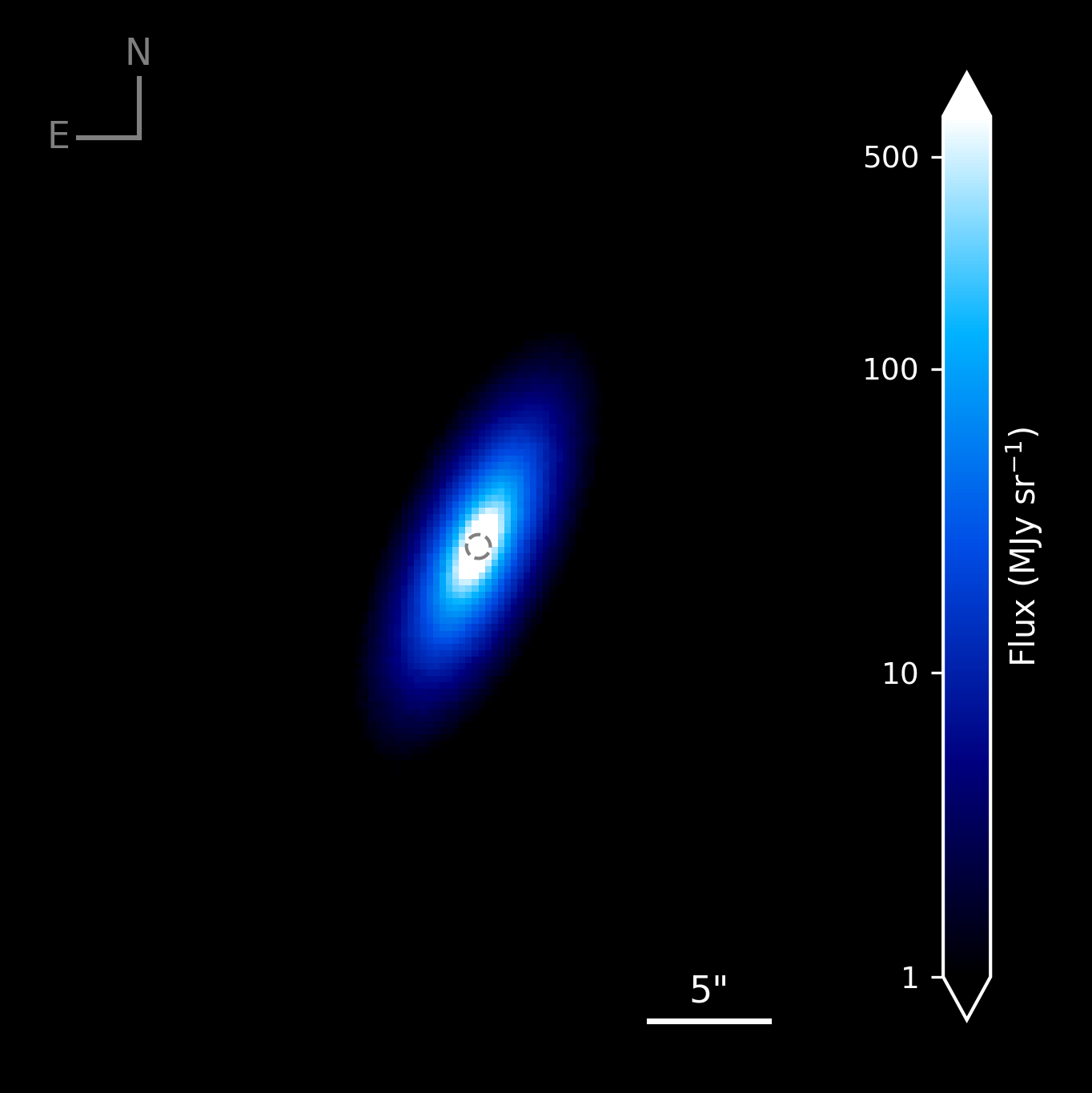}&
    \includegraphics[width=.3\textwidth,trim={0 0 0 0},clip]{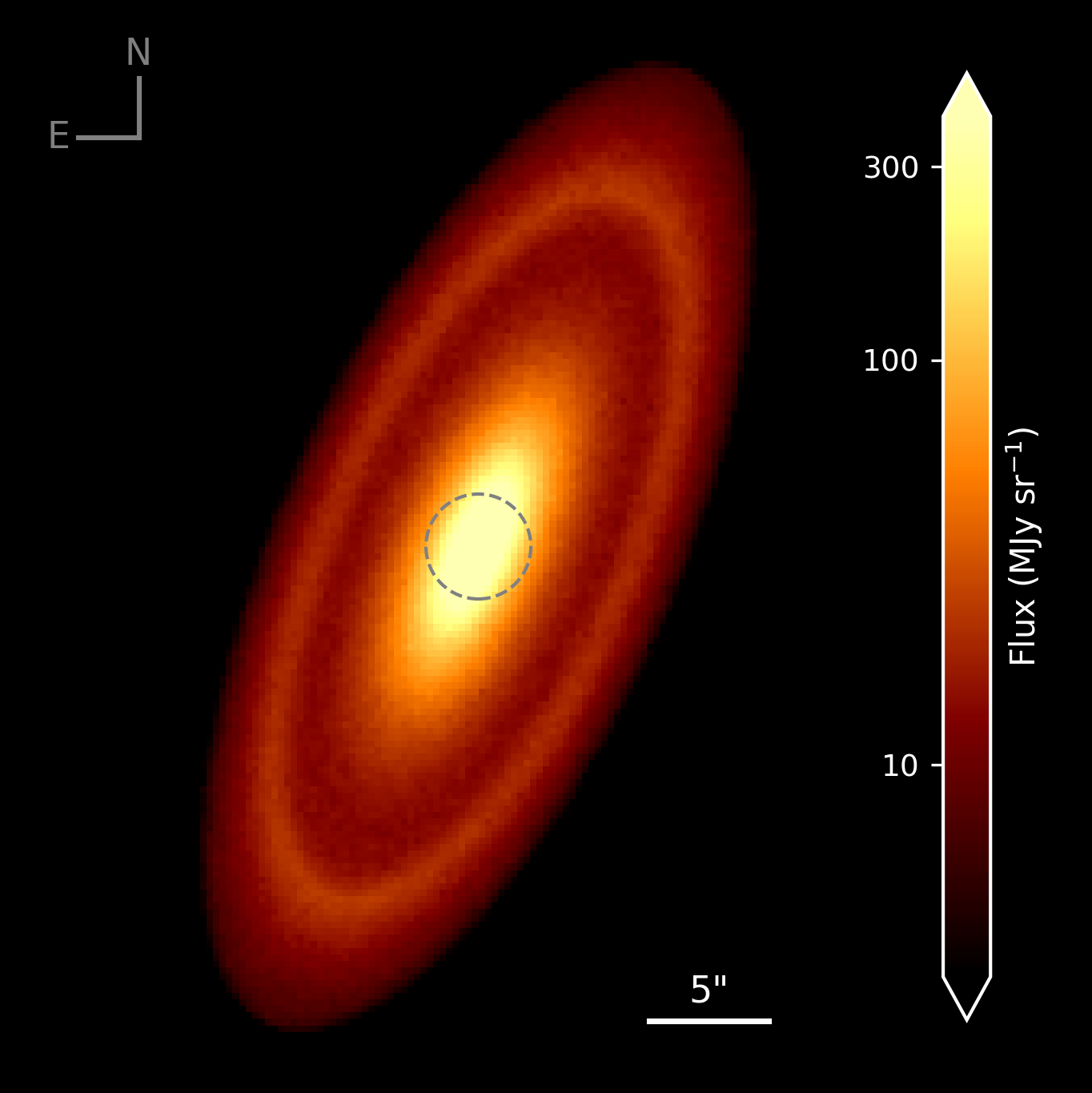}&
    \includegraphics[width=.3\textwidth,trim={0 0 0 0},clip]{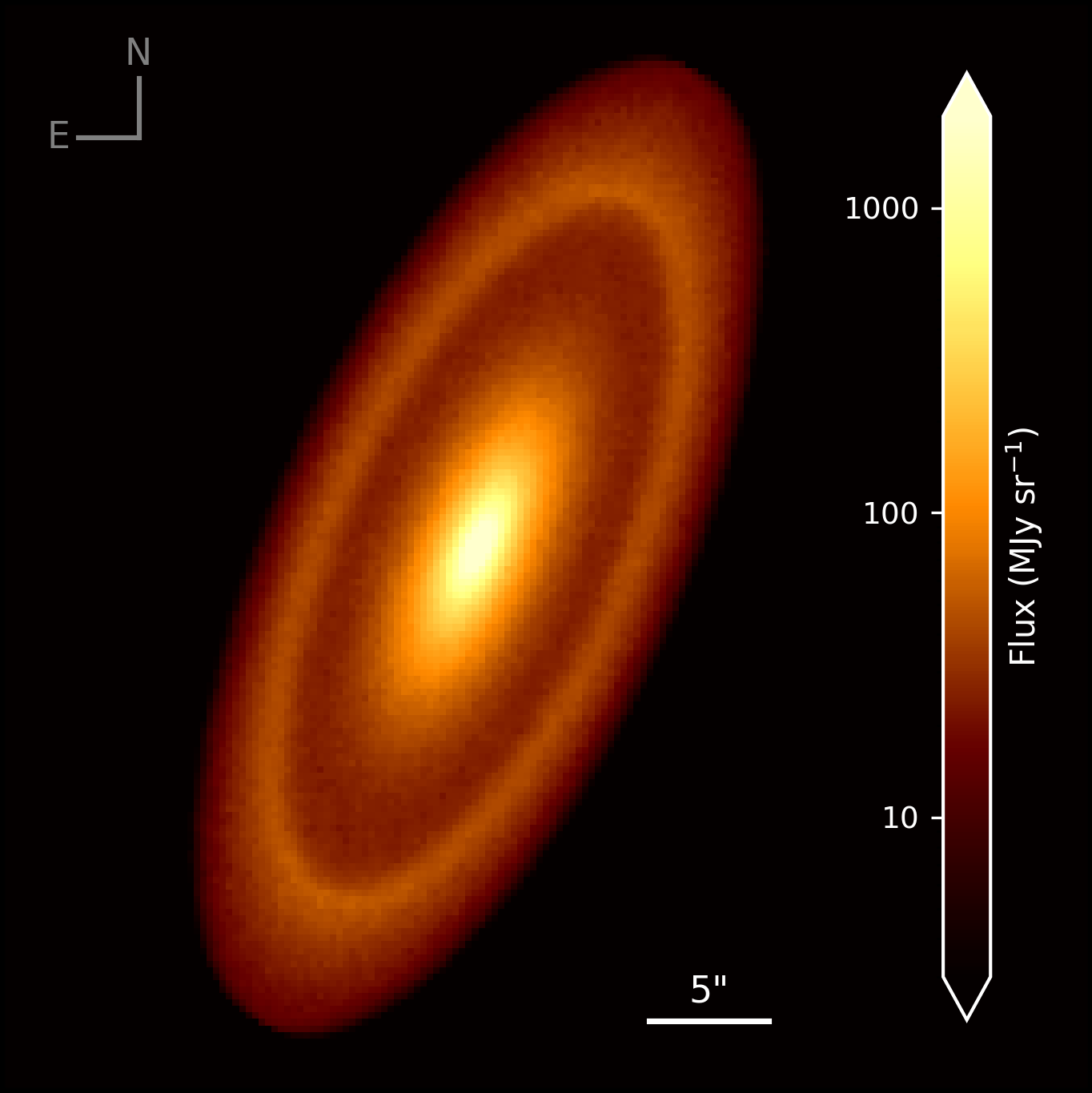}\\[.5ex]
    \hline\\[-1.4ex]
    \rotatebox{90}{\hspace{10mm} \Large after convolution}& &
    \includegraphics[width=.3\textwidth,trim={0 0 0 0},clip]{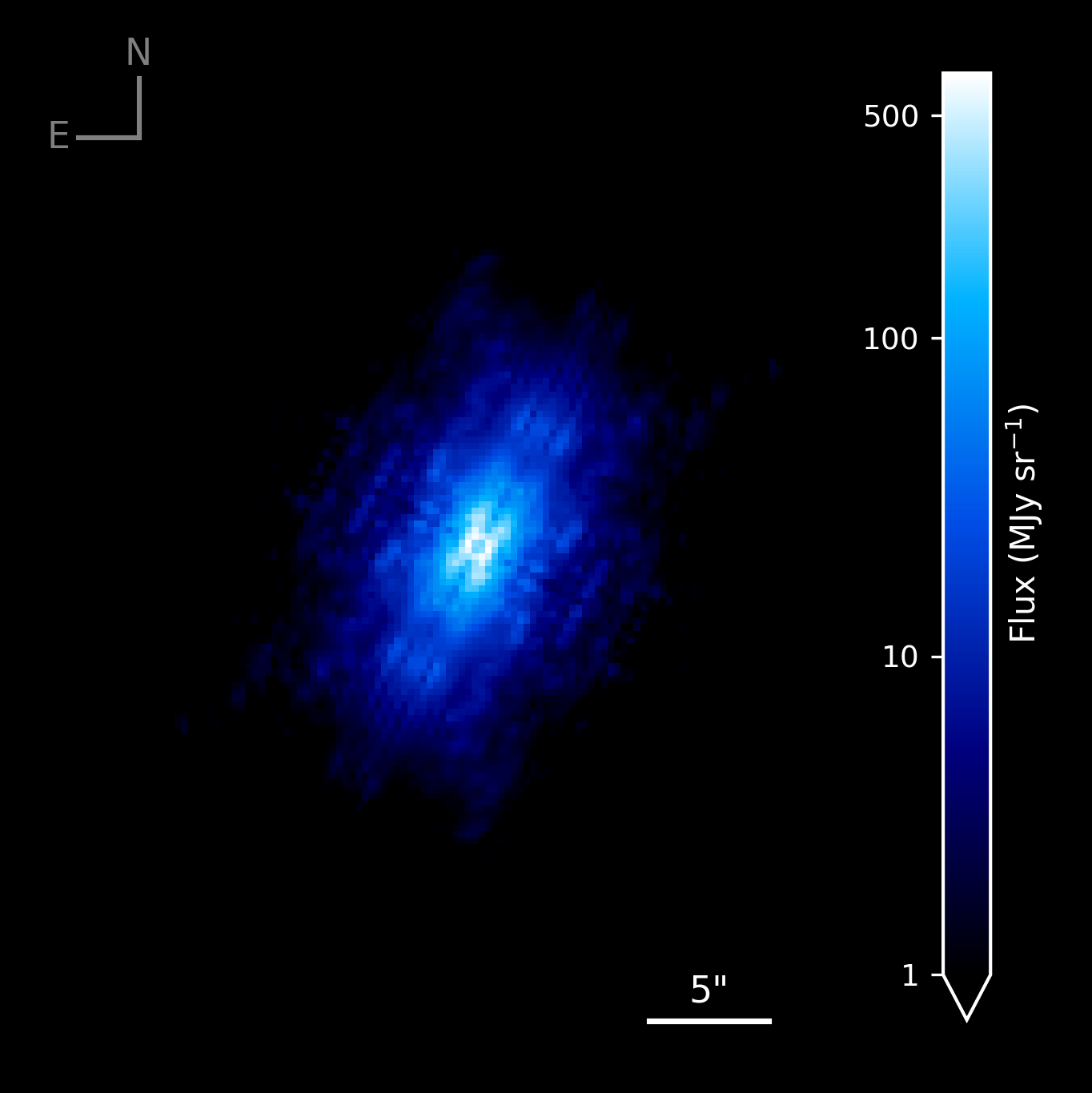}&
    \includegraphics[width=.3\textwidth,trim={0 0 0 0},clip]{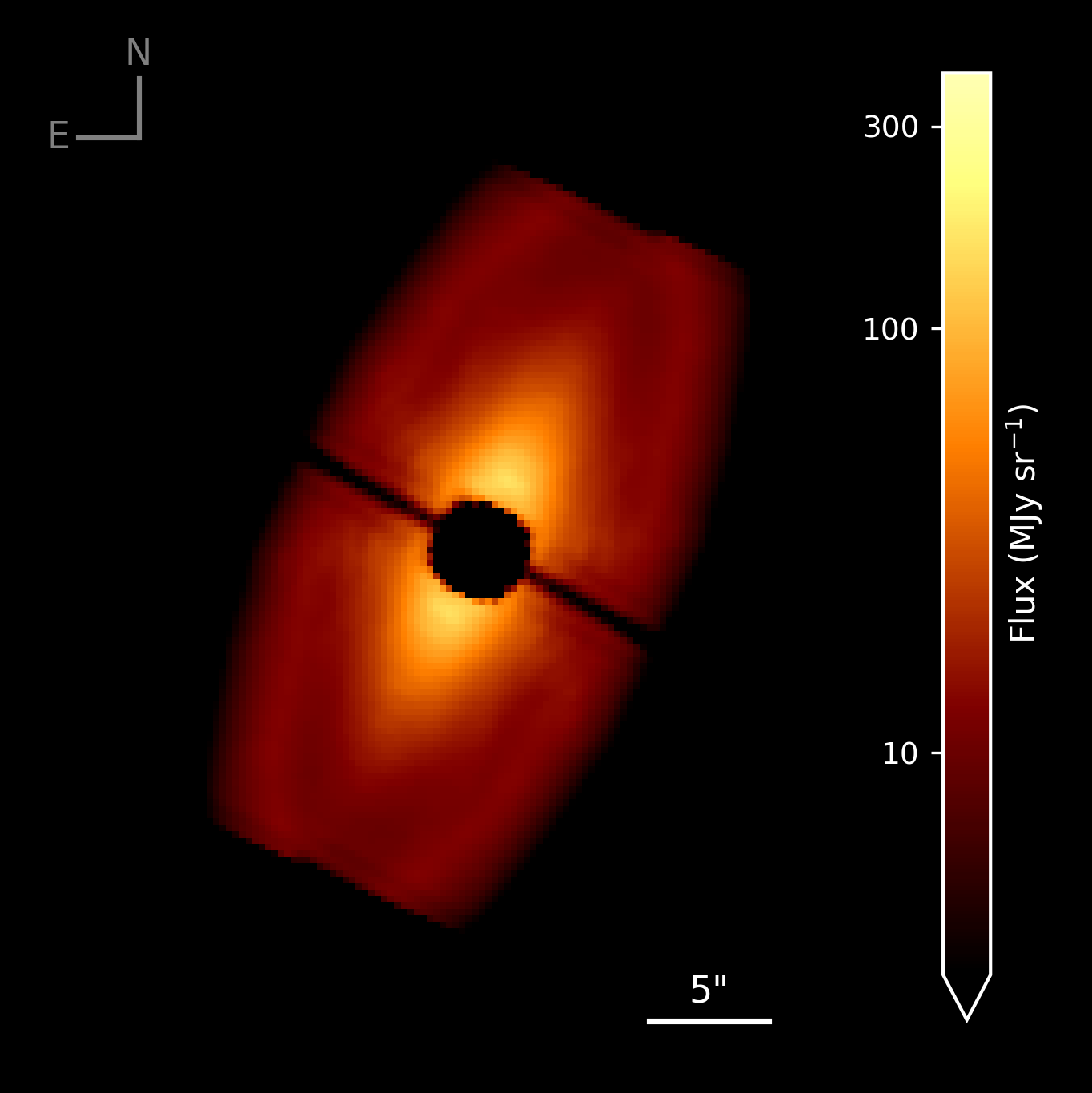}&
    \includegraphics[width=.3\textwidth,trim={0 0 0 0},clip]{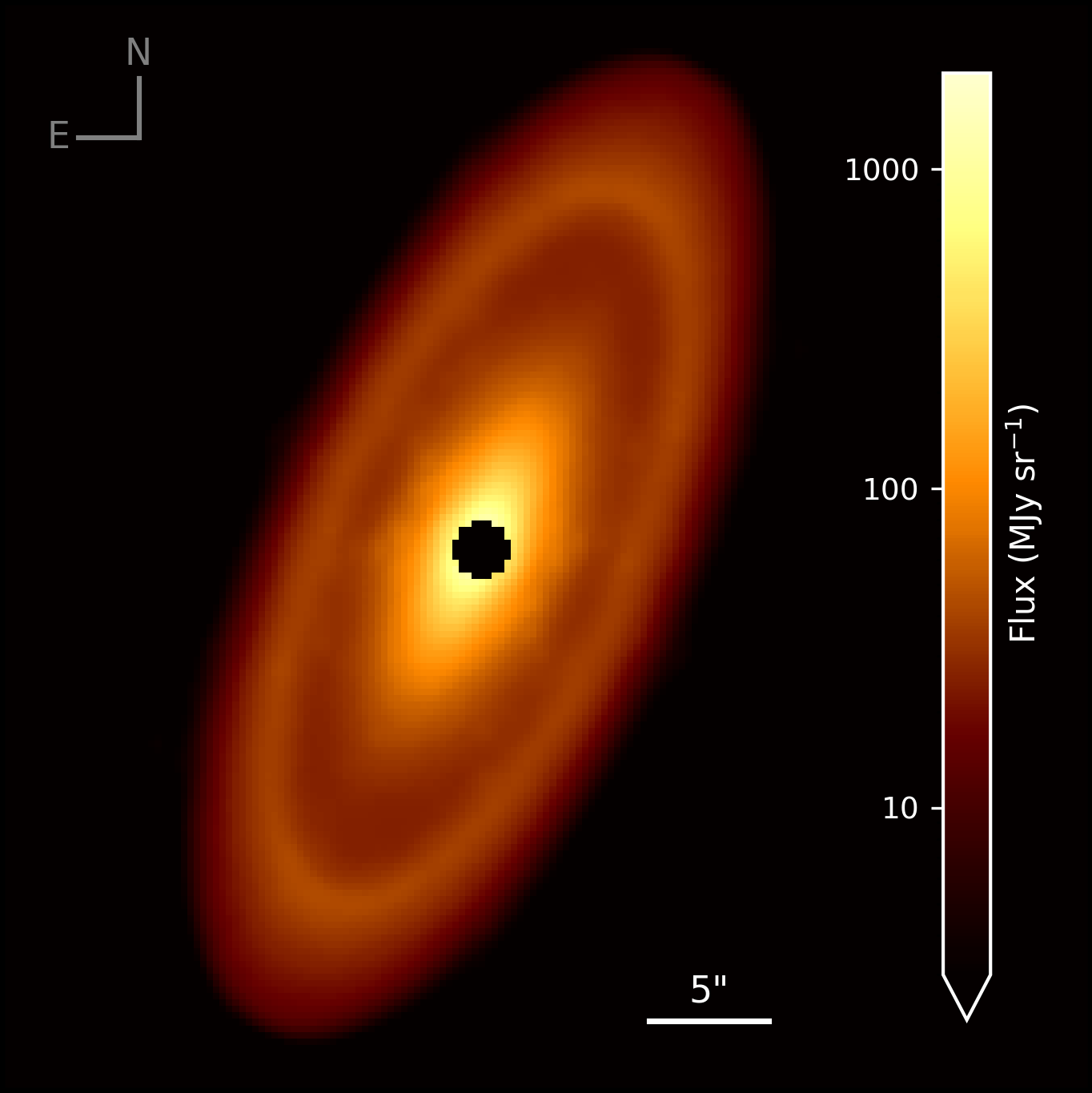}
    \end{tabular}
    \caption{Astrophysical scenes (top) and synthetic images (bottom) of the best-fit model disk,
    using the same colour scales as in \cref{fig:MIRI_images}.
    Dashed circles in the upper row indicate 
    the inner working angles of the coronagraphic filters.}
    \label{fig:best_fit_image}
\end{figure*}
}

In this section, we inspect the best-fit model outcome (Scenario A) more closely.
We first turn to the model's ability to reproduce the spatial distribution of emission
as observed by MIRI.
\cref{fig:best_fit_image} shows the synthetic images created from the best-fit model,
next to unconvolved images of the inclined disk, that is, the `pristine' astrophysical scenes
generated with \rave{} from the modelled face-on radial surface brightness profiles,
before the PSF of the respective filter is applied.
In the F2300C and F2550W images, we can easily discern the prominent outer belt, while
we note that a substantial amount of emission stems from the region within the belt.
In the case of the F1550C filter, only the hotter, innermost region of the disk is visible.
Comparing before/after PSF convolution, indicates a marked difference in the \qty{15.5}{\um} band,
where the original concentrated source is severely spread out to form a notable diffraction pattern, 
as previously described by \citet[][]{Boccaletti2024imaging}.

\begin{figure}
    \centering
    \includegraphics[width=\columnwidth,trim={0 0 0 0},clip]{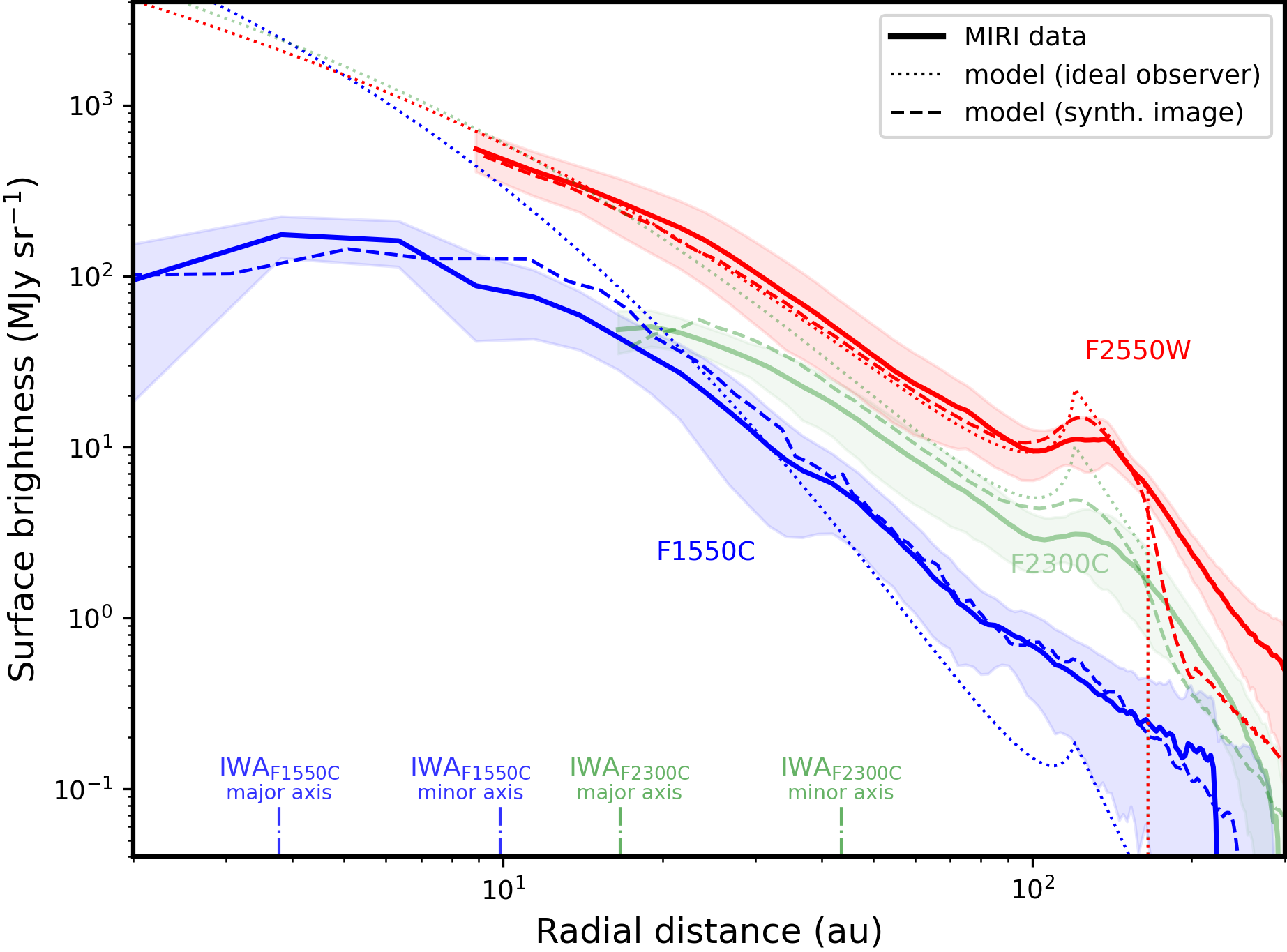}
    \caption{Radial brightness profiles---observation vs.\@ best-fit model.
    Solid lines represent profiles extracted from MIRI imaging
    \citep[image data from][]{Gaspar2023spatially}.
    \citep[Note that we reversed the additional calibration factor for the F2300C MIRI image applied by]
    [see text.]{Gaspar2023spatially}
    Shaded areas show the respective azimuthal 1$\sigma$ brightness variation at each radius.
    Dotted lines show the corresponding brightness profiles of the
    face-on and unconvolved best-fit model disk.
    Dashed lines show the brightness profiles retrieved from the synthetic images
    of the best-fit model.
    Also indicated are the distances corresponding to the IWAs of the coronagraphs
    on the disk major and minor axis.}
    \label{fig:best_fit_profiles}
\end{figure}

For the comparison with the MIRI images we resort to the radial brightness profiles,
shown in \cref{fig:best_fit_profiles}, which we obtain after deprojecting
the synthetic images in the same way as we do with the MIRI images.
We first note that the profiles derived from the synthetic images (dashed lines) differ substantially
from the pristine modelled profiles (dotted line), which correspond to an ideal, face-on observer 
(\ie{} without projection of the disk, PSF convolution, and subsequent deprojection).
In particular the steeper \qty{15.5}{\um} profile is considerably flattened,
again illustrating the blurring effect of the PSF already noticed in \cref{fig:best_fit_image}.
In the cases of the coronagraphs, this blurring is confounded by the drop-off in transmission
when approaching their inner working angles (IWA),\footnote{
    The IWA is the angular separation at which an off-axis point source 
    will have its transmission reduced to 50\%.
}
which correspond to radial distances of roughly 
\qty{4}{\astronomicalunit} and \qty{17}{\astronomicalunit} on the disk major axis, and
\qty{10}{\astronomicalunit} and \qty{43}{\astronomicalunit} on the disk minor axis
for the F1550C and F2300C filters, 
respectively \citep[$\mathrm{IWA_{F1550C}}\!=\!\qty{0.49}{\arcsecond}$
and $\mathrm{IWA_{F2300C}}\!=\!\qty{2.16}{\arcsecond}$,][]{Boccaletti2015midinfrared}.
These distances are also indicated in \cref{fig:best_fit_profiles}.
Note that, while its IWA might be smaller, the transmission drop-off of the 4QPM coronagraph,
F1550C, extends relatively further out,
owing to its tapered transmission profile \citep[with $\sim\!66\%$ transmission
at $2\!\cdot\!\mathrm{IWA_{F1550C}}$,][Fig.~6 therein]{Boccaletti2015midinfrared}.

Comparing the synthetic image profiles (dashed lines) to those retrieved from actual MIRI images
(solid lines) we find considerable agreement.
Significant deviations only appear exterior to the planetesimal belt (\qty{>168}{\astronomicalunit}),
where the model brightness drops off more steeply than was observed.
This is explained by the fact that our simplified model does not include
a representation of the disk halo, that is, the ensemble of small dust near and below the blow-out size
that is pushed by radiation pressure beyond the outer belt.
Nevertheless, due to the PSF blurring effect combined with the inclination of the disk,
considerable brightness even beyond the belt---where no dust exists in the model---is 
present in our synthetic images and derived radial profiles,
forming a faint pseudo-halo.
Conversely, we can infer that part of the halo observed by MIRI
is PSF-blurred emission from the planetesimal belt, rather than actual halo dust.

Note that the model was only fitted to the profiles retrieved from the 
F1550C and F2550W images, given the issues with the F2300C calibration
pipeline suspected by \citet{Gaspar2023spatially}.
Nevertheless, we have plotted here the radial profile retrieved from the F2300C image,
however, only after reversing the additional, `corrective' scaling factor of \num{3.14}
that \citet{Gaspar2023spatially} had applied in order to align the F2300C and F2550W fluxes.
The consistent offset of a factor \num{\sim3} between the resolved F2300C and F2550W fluxes
in their original reduction (apparent in the gap between the solid lines in \cref{fig:best_fit_profiles})
has led \citet{Gaspar2023spatially} to surmise (quite reasonably) an issue in the calibration pipeline, 
given that \textit{Spitzer} spectrographic observation of the inner disk had revealed a nearly flat spectrum
at those wavelengths, with only a few percentage points increase 
from \qty{23}{\um} to \qty{25.5}{\um}.
(We will discuss the \textit{Spitzer} spectral data in more detail in \cref{sect:irs}.)
However, we see here that, while the underlying surface brightness (dotted line) at those wavelengths
is indeed very similar at shorter radial distances,
the corresponding brightness profiles retrieved from the synthetic images (dashed lines) 
differ consistently by a factor of \numrange{2}{3},
owing in particular to the aforementioned transmission drop-off at short working angles.
We therefore conclude that there is no evidence for an inherent issue with the F2300C reduction pipeline.

\begin{figure}  
    \centering
    \includegraphics[width=\columnwidth]{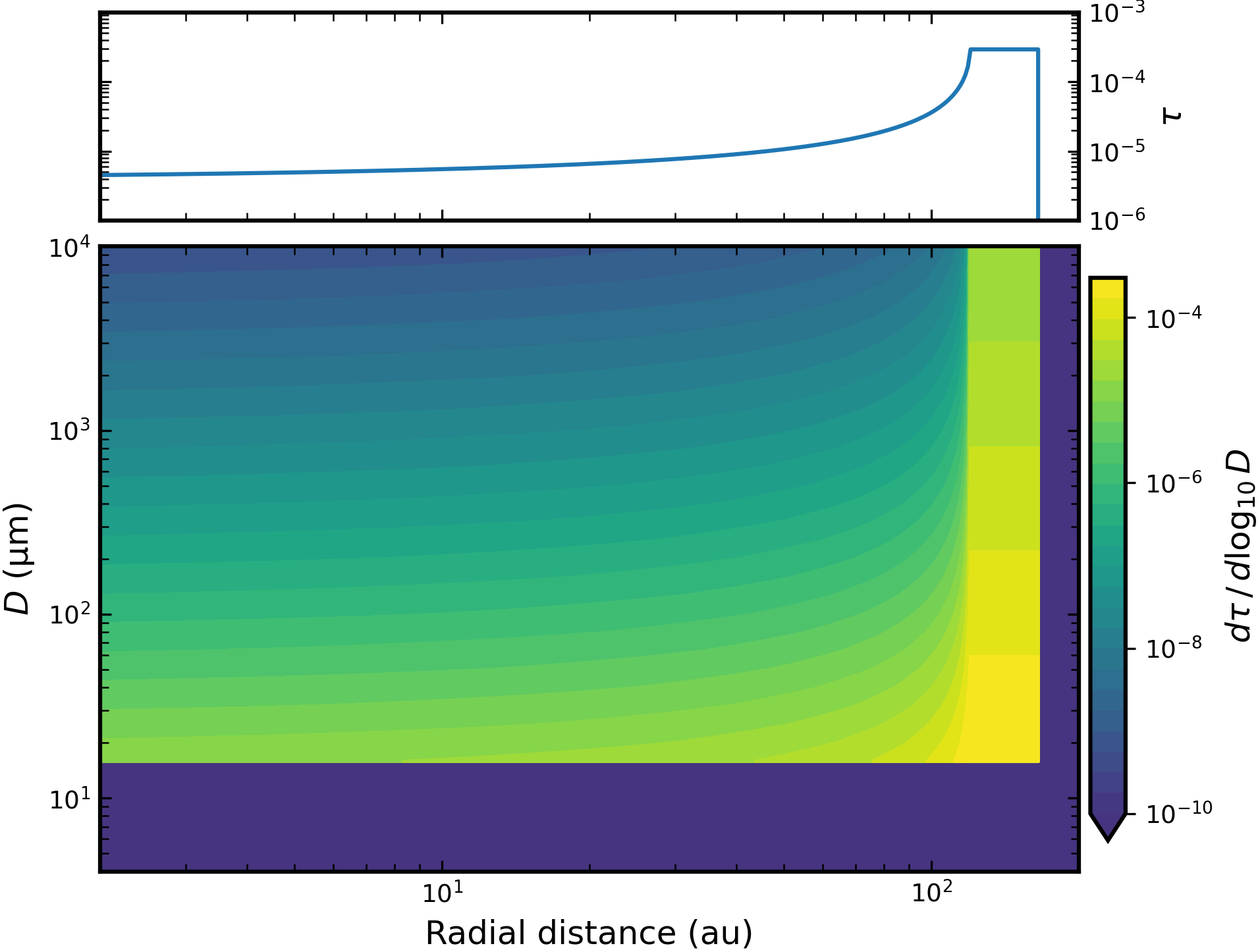}
    \caption{Radial distribution of dust of the best-fit model.
    Top: Face-on optical depth over radius, integrated over all particle sizes.
    Bottom: Face-on optical depth over radius and particle size.
    The colour scale gives the optical depth per unit size decade.}
    \label{fig:best_fit_tau}
\end{figure}

On the whole, we conclude that the modelled brightness distribution
matches the one observed by MIRI remarkably well. 
The underlying radial surface density distribution of the model disk is illustrated in \cref{fig:best_fit_tau}.

\begin{figure}
    \centering
    \includegraphics[width=\columnwidth,trim={0 0 0 0},clip]{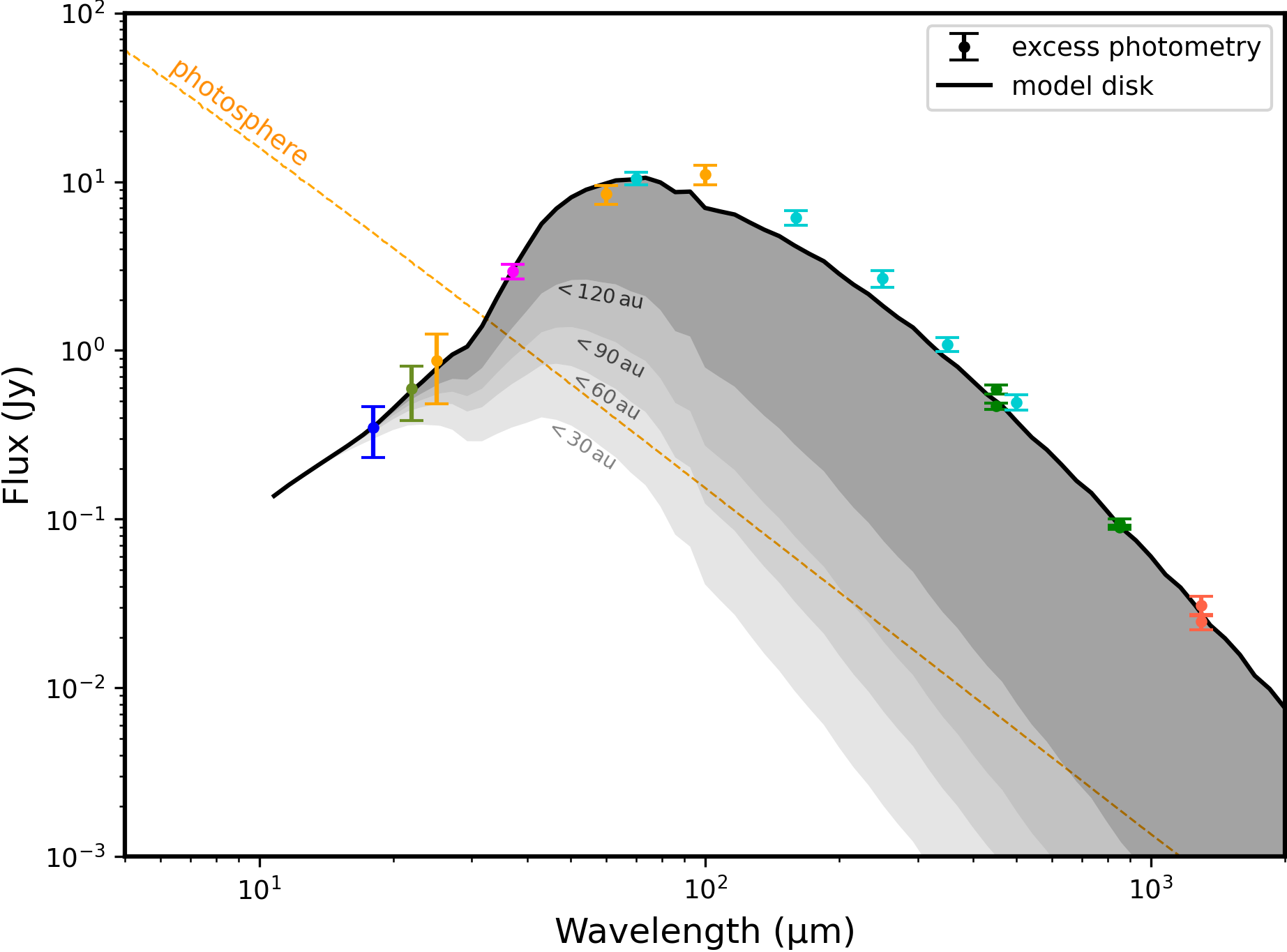}
    \caption{SED of the best-fit model disk compared to the excess photometry 
    (\cref{tab:phot_data}; marker colours as in \cref{fig:SED_phoenix}).
    Shaded areas indicate the contribution to the SED from different regions of the disk, 
    each corresponding to a specific maximum disk radius.
    The photosphere model flux is given for reference.}
    \label{fig:best_fit_SED}
\end{figure}

Secondly, we inspect disk's simulated SED, which is given in \cref{fig:best_fit_SED}.
Overall, we find that the model SED aligns closely with the observed excess datapoints,
which the model was fitted to.
Only in the regime of \qtyrange{100}{300}{\um} there is a consistent
underprediction of the observed fluxes, with deviations up to $\sigma\!=\!3.3$,
which, given the simplifications of the model, we deem acceptable.
We thus note that the total disk emission is reasonably well reproduced by the best-fit model.
\cref{fig:best_fit_SED} also shows the contribution to the total emission 
from different (radial) parts of the disk.
Up to the wavelength of \qty{18}{\um}, almost all emission comes from regions within \qty{30}{\astronomicalunit}.
Beyond \qty{18}{\um}, the emission from within \qty{30}{\astronomicalunit} stagnates,
while contributions from the outer regions significantly rise.
At \qty{33}{\um}, the outer belt (regions \qty{>120}{\astronomicalunit}) becomes the dominant contributor
with 50\% of the total emission, rising to 70\% at \qty{50}{\um},
where emissions from the inner regions level off, and 90\% at \qty{100}{\um}.
At \qty{1.3}{\mm}, the wavelength of the ALMA datapoints, regions within \qty{120}{\astronomicalunit}
contribute only \num{1.1}\% to the total flux.
(Note that the ALMA datapoints represent only the direct flux from the belt,
which due to the low contribution from the interior regions still align
with the total emission SED shown in \cref{fig:best_fit_SED}.)
The face-on surface brightness generated by our model at \qty{1.3}{\mm} interior to the belt
is minuscule, due to a dearth of large particles, and within \qty{100}{\astronomicalunit} 
relatively constant at around \qty{.01}{\mega\jansky\per\steradian} 
($\approx\!\qty{2e-4}{\milli\jansky\per\square\arcsecword}$).

  \section{Comparison with IRS spectrum} \label{sect:irs} 

For a more detailed examination around mid-infrared wavelengths
we additionally compare the best-fit model to the spectra taken by the Spitzer Infrared Spectrograph (IRS),
namely the high-resolution spectra taken of the central region
\citep[PID~90, reduction by][]{Su2013asteroid} and the low-resolution spectrum
taken of the south-east ansa of the outer ring
\citep[PID~1074, reduction from the Spitzer Heritage Archive, see also][]{Stapelfeldt2004first}.
Given the complexity of how the disk emission contributes to the spectrum (as outlined below)
we chose not to include this data in the fitting and rather
to perform a consistency check on the best-fit model.

\begin{figure}
    \centering
    \includegraphics[width=\columnwidth,trim={0 0 0 0},clip]{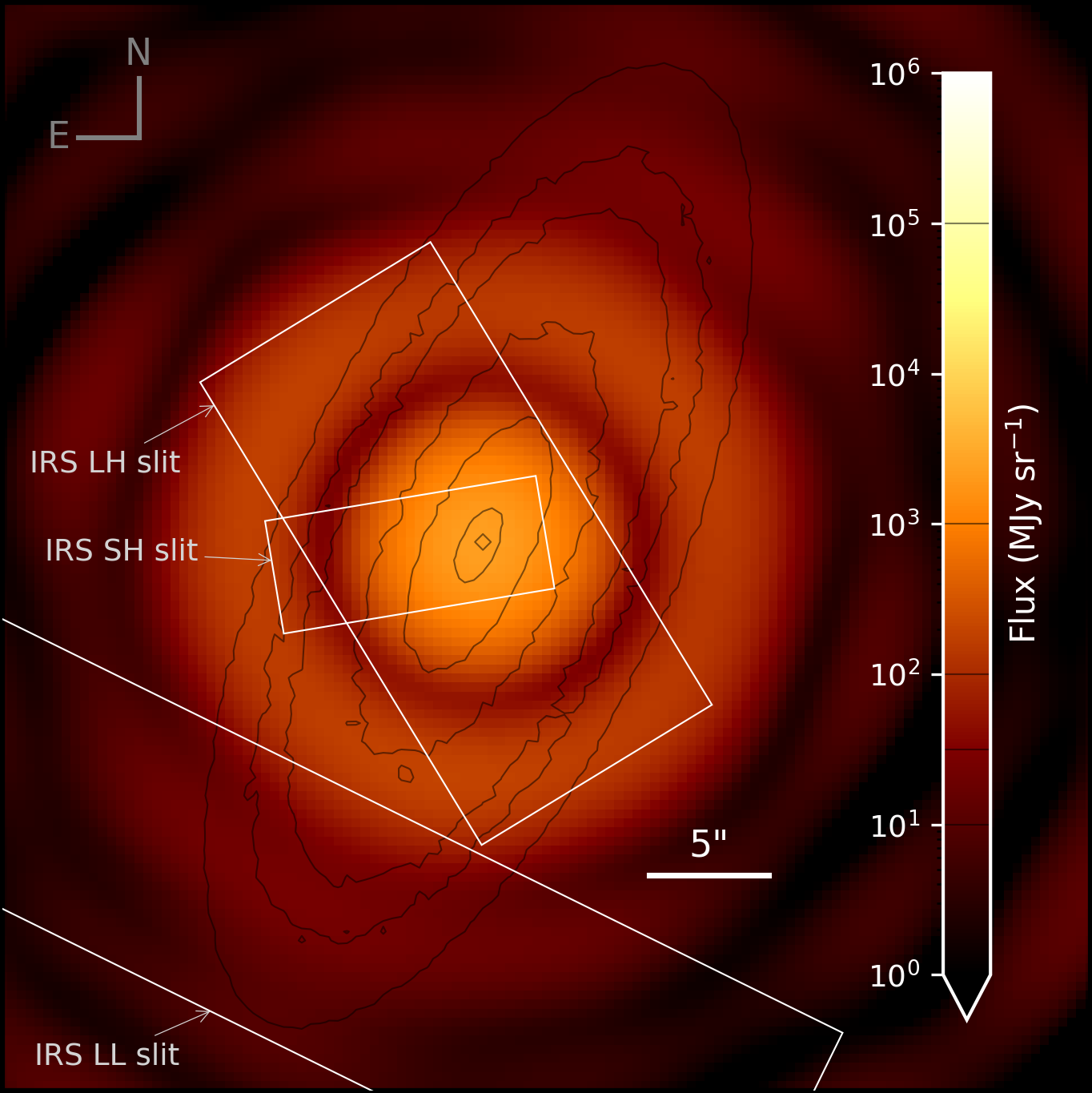}
    \caption{IRS module slits used for spectrum extraction overplotted
    onto an exemplary pseudo-image of the disk and star at a wavelength of \qty{25}{\um}
    (after convolution with the corresponding LH module PSF from \tinytim).
    Also shown are contours of the unconvolved astrophysical scene 
    to indicate the location of the disk/belt and the star
    (levels shown on colourbar; stellar flux distributed onto the central pixel).
    The PID~90 `on-star' spectrum is measured in the SH (\nth{2} nod position) 
    and LH (central mapping position) slits at the respective wavelengths.
    The PID~1074 `on-ansa' spectrum is measured in the displayed LL slit tapered aperture.}
    \label{fig:IRS_slits}
\end{figure}

\begin{figure}
    \centering
    \includegraphics[width=\columnwidth,trim={0 0 0 0},clip]{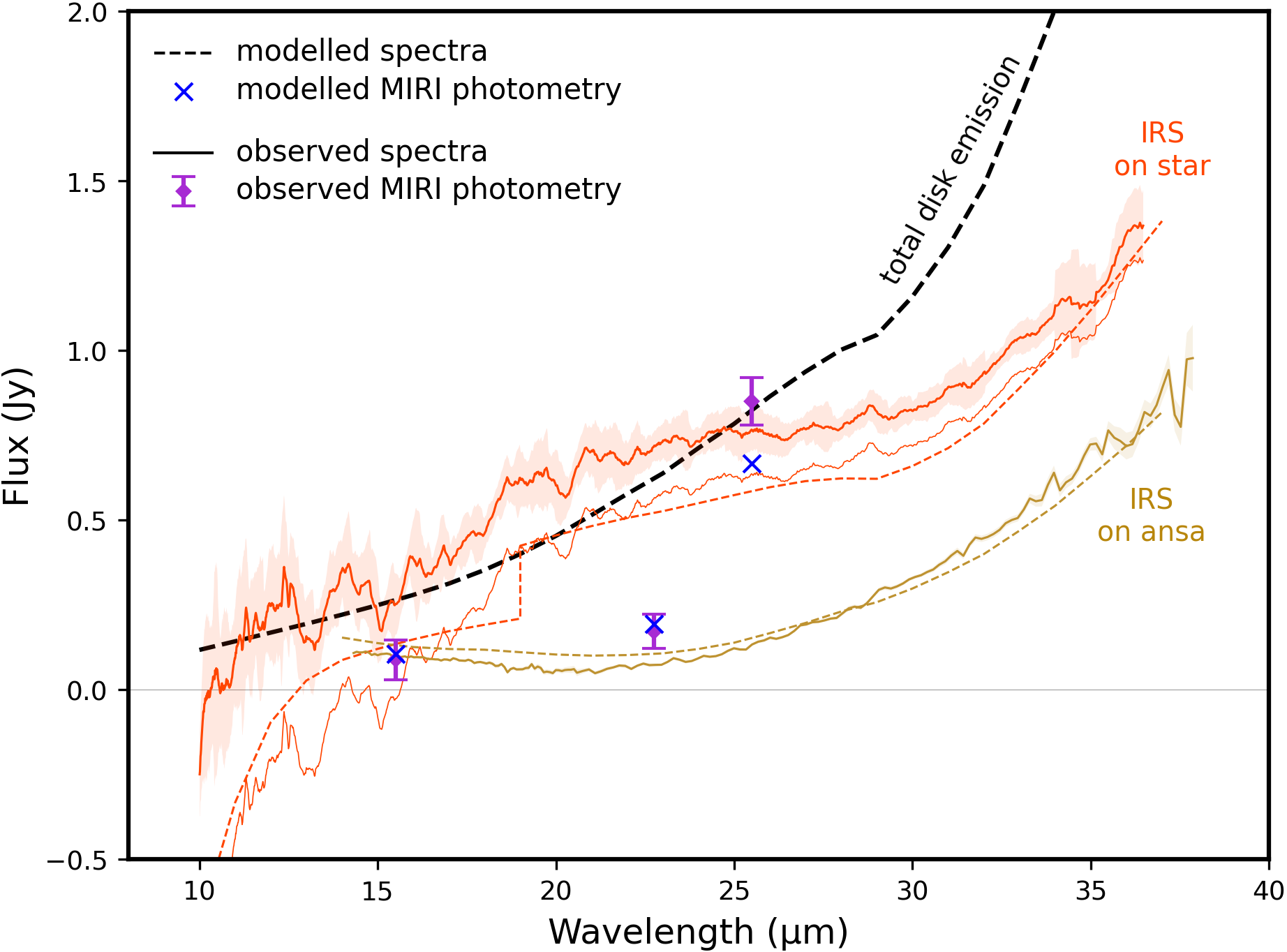}
    \caption{Simulated IRS spectra (dashed coloured lines) and MIRI photometry 
    (\ie{} total image fluxes, blue crosses) of the best-fit model disk.
    Note that these do not align and also differ from the total emission of the underlying disk 
    (black dashed line) due to instrumental effects and viewing geometries, see text.
    These simulated spectra and photometry are instead to be compared with the corresponding
    observed spectra (solid coloured lines with error corridors) 
    and observed MIRI photometry (diamonds with error bars) \citep[from][Fig.~3 therein; 
    note that we have reversed their additional calibration factor of 3.14,
    see text]{Gaspar2023spatially}.
    We show the IRS on-star high-resolution spectrum from PID~90
    \citep[reduction by][with our model photosphere flux subtracted]{Su2013asteroid}
    as well as the IRS on-ansa low-resolution spectrum from PID~1074 (reduction from SHA).
    Also shown is a version of the observed on-star spectrum, in which the total fluxes were reduced 
    by 4\% prior to photosphere subtraction; shown as the thin solid line.}
    \label{fig:best_fit_SED_closeup}
\end{figure}

To simulate IRS spectra taken of our model disk, we generate pseudo-images
of the disk and central star at a range of wavelengths (at \qty{1}{\um} steps) 
and measure the total flux within the extraction field of the respective spectrum recording.
The pseudo-images are synthesized by first generating astrophysical scenes of the inclined disk (with \rave{})
with a superimposed star represented by a point source rescaled to the stellar flux at each wavelength.
These scenes are then convolved with the respective PSFs for each IRS module at each wavelength,
which are obtained from the \tinytim{} tool \citep{Krist2006tinytim}.
An example of such a pseudo-image with outlines of the corresponding extraction fields
is given in \cref{fig:IRS_slits}.
The different IRS modules are designated SH (shorter wavelengths, high resolution) and
LH (longer wavelengths, high resolution) used for the on-star spectrum of PID~90,
and LL (longer wavelengths, low resolution) used for the on-ansa spectrum of PID~1074.

We determine the relative positioning of the extraction fields using the pointing information
from the corresponding observation files on the Spitzer Heritage Archive (SHA), 
combined with Fomalhaut's position, calculated at the time of the observations 
using coordinates and proper motions provided by the \textit{Hipparcos} mission 
\citep[][data obtained from SIMBAD database]{Leeuwen2007validation}.

To mimic the reduction of the reference spectra, we also apply the slit loss correction
for a point source to the simulated IRS spectra, even though
the underlying sources are not strictly point sources.
The corresponding correction factors were calculated as the inverse of the fraction
of light that falls within the slit from a point source at the slit centre
(or at the nod reference locations at 1/3 or 2/3 of the slit length in the case of the SH module)
convolved with the respective \tinytim{} PSF.
For the high-resolution spectrum, we merge the partial spectra of the SH and LH modules,
computed individually up to, and down to \qty{19}{\um}, respectively, 
where their wavelength ranges overlap.
Note that we did not mimic the wavelength-dependent variable width extraction of the LL module extraction,
but simply measured the total flux within the tapered slit aperture (seen in \cref{fig:IRS_slits})
resulting from the two-nod position reduction pipeline, which may affect their comparability.
On the other hand, the full-slit extraction of the SH \& LH module reduction pipeline
should be comparable to our approach.

The resulting simulated spectra are shown alongside the observed ones in \cref{fig:best_fit_SED_closeup}.
In the case of the on-star spectrum from PID~90, we compare the excesses
(\ie{} we subtract our model photosphere fluxes from both the simulated and the observed spectrum)
whereas the on-ansa spectrum from PID~1074 combines the local disk emission with the
PSF wings of the central source.
First off, it should be noted that the on-star spectra (both modelled and observed) 
consistently diverge from the total emission of the model disk at longer wavelengths, 
as expected, due to the only partial coverage of the cold outer ring by the slit field of view.
The observed spectra are, naturally, more appropriately compared to our simulated IRS spectra.

For the on-star spectrum, we find that fluxes are underpredicted by our model,
although the spectrum's overall shape appears to be reasonably reproduced.
In particular the marginal slope at wavelengths of \qtyrange{20}{30}{\um},
which markedly diverges from the trend in the total disk emission, is reproduced well.
This flattening of the spectrum is caused by the size and orientation of
the LH module's extraction field of view.
At these wavelengths, the cold outer belt rapidly gains in brightness, 
which is largely missed by the LH module field of view.
Tests with slit orientations aligned with the disk major axis,
which better capture the outer belt, resulted in a spectrum that is more aligned
with the steeper curve of the model disk's total emission.

Moreover, the discontinuity at \qty{19}{\um} in our simulated spectrum appears to resemble
sudden flux increase in the observed spectrum at around \qtyrange{18}{19}{\um}.
This discontinuity is due to our discrete switchover from the SH to the LH module,
which, given their different fields of view, are differently affected by
the not strictly applicable point source correction and the imperfect pointing.
For one, the point source slit loss correction can overcompensate for the somewhat---and
at longer wavelengths increasingly so---extended source.
(Note that for an entirely flat source, slit losses are exactly compensated
by gains from off-slit sources making no correction necessary.)
Compared to the SH slit, the wider LH slit captures more of that extended-source flux,
which is also amplified by the point source correction,
allowing the simulated spectrum to even slightly exceed the actual total disk flux
at around \qtyrange{19}{20}{\um}.
On the other hand, the offset of the slit centre (or the nod reference positions)
from the stellar position can cause the slit loss correction to undercompensate.
While the offset of the reference positions from the star is similar in the nodded SH
and the LH observations (\qtyrange{\sim.5}{1}{\arcsecond}), this effect is more
significant in the case of the smaller field of view of the SH module,
and even causes the simulated excess spectrum to become negative towards smaller wavelengths.
We find that at \qty{<15}{\um}, where the disk flux is on the order of 1\% of the stellar flux,
slit position changes of even just a few \qty{100}{\milliarcsec} severely impact
the resulting simulated IRS excess spectrum.

The overall offset between the observed and the simulated on-star excess spectra 
may be reconciled by considering uncertainties in the absolute calibration of the total fluxes. 
With the stellar photosphere flux being the dominant contributor over the IRS wavelength range,
even minor differences in the calibrated total fluxes (or the level of photosphere
subtracted) lead to significant differences in the excess fluxes.
\citet{Su2013asteroid} estimate a spectrophotometric accuracy of 5\% of their reduction
which includes a calibration with custom spectral response functions derived from spectra of other stars
(which we do not replicate here).
The observed excess spectrum can be brought into reasonable alignment with the simulated one
by reducing its total fluxes by \num{\sim4}\%, as demonstrated in \cref{fig:best_fit_SED_closeup}.
In fact, a re-calibration of the spectrum of \citet{Su2013asteroid} by \citet{Gaspar2023spatially}
has brought about a flux reduction of that magnitude, 
which would improve the fit of our simulated spectrum.

In the case of the on-ansa spectrum, we find that our model reproduces the overall spectrum shape well,
yet at shorter wavelengths, it overpredicts the fluxes by a factor of up to two (at around \qty{20}{\um}).
This is to some extent surprising. 
If anything, we would rather have expected an underprediction by our model,
since the south-east ansa of the actual Fomalhaut disk appears somewhat brighter
than the north-west ansa \citep{Gaspar2023spatially}---an asymmetry
which by construction we do not intend to reproduce given our axisymmetric model.
However, at these shorter wavelengths, the spectrum is dominated by the PSF wings
of the stellar source, not by the cold outer belt, which also gives rise to the
well-reproduced negative slope \citep[as also noted by][]{Stapelfeldt2004first}.
This suggests that the discrepancy more likely stems from inaccuracies in our 
simplified reduction or in the calibration of the observed spectrum,
rather than from differences in the underlying disk emission.

We may also compare the IRS fluxes to the total fluxes in the MIRI images, 
observed and synthetic ones, which are overplotted in \cref{fig:best_fit_SED_closeup}. 
It is evident that the total fluxes in the synthetic coronagraphic images, F1550C and F2300C, 
align neither with the on-star IRS spectrum, nor with the actual total emission of the model disk.
This is because much of the flux originating from within several 10s of \unit{\astronomicalunit} 
is lost due to the transmission drop-off at short working angles,
as already noted in \cref{sect:best_fit}.
Additional losses occur due to the limited field of view of these filters,
particularly in the case of the Lyot coronagraph, F2300C,
where emission from the outer belt is significant.
The total fluxes in the coronagraphic images are thus only 36\% (F1550C)
and 30\% (F2300C) of the model disk's total emission.
In contrast, 79\% of the flux is recovered in the non-coronagraphic image,
F2550W, all losses (\qty{157}{\milli\jansky}) being due to our masking of pixels within 
\qty{1}{\arcsecond} from the star to mimic the saturation in the observed image.

We find that, in the case of the coronagraphic filters,
the total image fluxes in the synthetic images align well
with those from \citet{Gaspar2023spatially} (after reversing their additional 
scaling factor of \num{3.14} for the F2300C image, see \cref{sect:best_fit}).
The underprediction of the F2550W image flux by about \qty{200}{\milli\jansky} 
stems from our model's aforementioned omission of the disk halo,
which was captured by the wider field of view of this filter.
Including such a halo in our model would increase the total disk flux 
and the total flux in the synthetic F2550W image accordingly, 
while leaving observables with more constrained fields of view, 
such as the coronagraphic images and the on-star IRS spectra, relatively unaffected.

Given that we fitted our model to the F1550C and F2550W radial profiles up to the outer
edge of the belt, the agreement (F1550C) and underprediction (F2550W) 
of total image fluxes are anticipated.
The consistently low F2300C total fluxes, on the other hand, are reassuring
and underscore the challenge of comparing such coronagraphic photometry
with the IRS on-star excess spectrum.

  \section{Discussion} \label{Sect:Discn}

This study sought to determine whether Poynting-Robertson (PR) drag transport
from the outer belt alone could explain the significant amount of dust 
observed interior to the outer planetesimal belt of the Fomalhaut debris disk. 
To do this, we applied an analytical PR drag model to the Fomalhaut system,
which allowed the exploration of a vast particle parameter space,
pertaining to the dust composition and collisional strength.
In a parameter grid search, the model was simultaneously fitted to the system's excess emission SED
and to radial brightness profiles extracted from resolved JWST/MIRI images.

We found that the observed radial characteristics of the extended disk could be explained 
solely by PR-drag transport, particularly when considering instrumental effects, 
namely PSF blurring and the transmission drop-off of the coronagraphs.
Good fits to the observations postulate a significant water ice component in the dust grain material,
and place firm constraints on the collisional strength of particles in the 10s of microns size regime.

Further validation of the best-fit model is given by comparison with IRS spectra.
Our simulated spectra reproduce the general features of the observations, in particular 
the flatness of the star-centred spectrum at \qtyrange{20}{30}{\um}, which we find
is largely due to the orientation of the extraction aperture of that particular IRS observation, 
not necessarily reflecting the trends in the disk's total emission.

The degree to which our rather simplistic model can reproduce the observations
may appear surprising.
However, the model's ability to consistently explain various types of data---that is,
the SED, MIRI-retrieved radial brightness profiles, and IRS spectra---suggests
robustness rather than mere adaptation to specific dataset anomalies.
In fact, the fair reproduction of the F2300C profile and the IRS spectra
occurred without them being fitted to.
This robustness is supported by the limited number of free parameters and the analytical nature of the PR drag model,
being applied broadly without tailoring excessively to any single feature of the Fomalhaut system.

Significant simplifications inherent to the model include: the absence of planets affecting the disk structure
\citep[\eg{} through ejection or resonant trapping,][]{Bonsor2018using,Sommer2020effects};
the omission of other dust sources, such as comets or another collisional planetesimal reservoir 
\citep[\eg{} in the region of the warm inner disk component, 
termed `asteroid belt analogue' by][]{Gaspar2023spatially};
and the assumption that a single material composition 
and collisional strength prescription apply throughout the disk.
Nonetheless, given the level of detail we are aiming to emulate---which 
notably excludes azimuthal features like the intermediate ring---the
Fomalhaut disk may not require a more complex representation.
Specifically, it might lack phenomena we have not modelled, 
such as giant planets capable of carving distinct cavities or comets
delivering substantial dust quantities.
In any case, our modelling supports the notion that the dust distribution 
in the Fomalhaut disk is consistent with a pure PR drag transport scenario.
We will explore the implications of some of these simplifications further in the following sections,
where we discuss aspects of the model and our obtained parameter constraints in more detail.

\subsection{Collisional strength} \label{Sect:collstr}
\begin{figure} 
    \centering
    \includegraphics[width=\columnwidth,trim={0 0 0 0},clip]{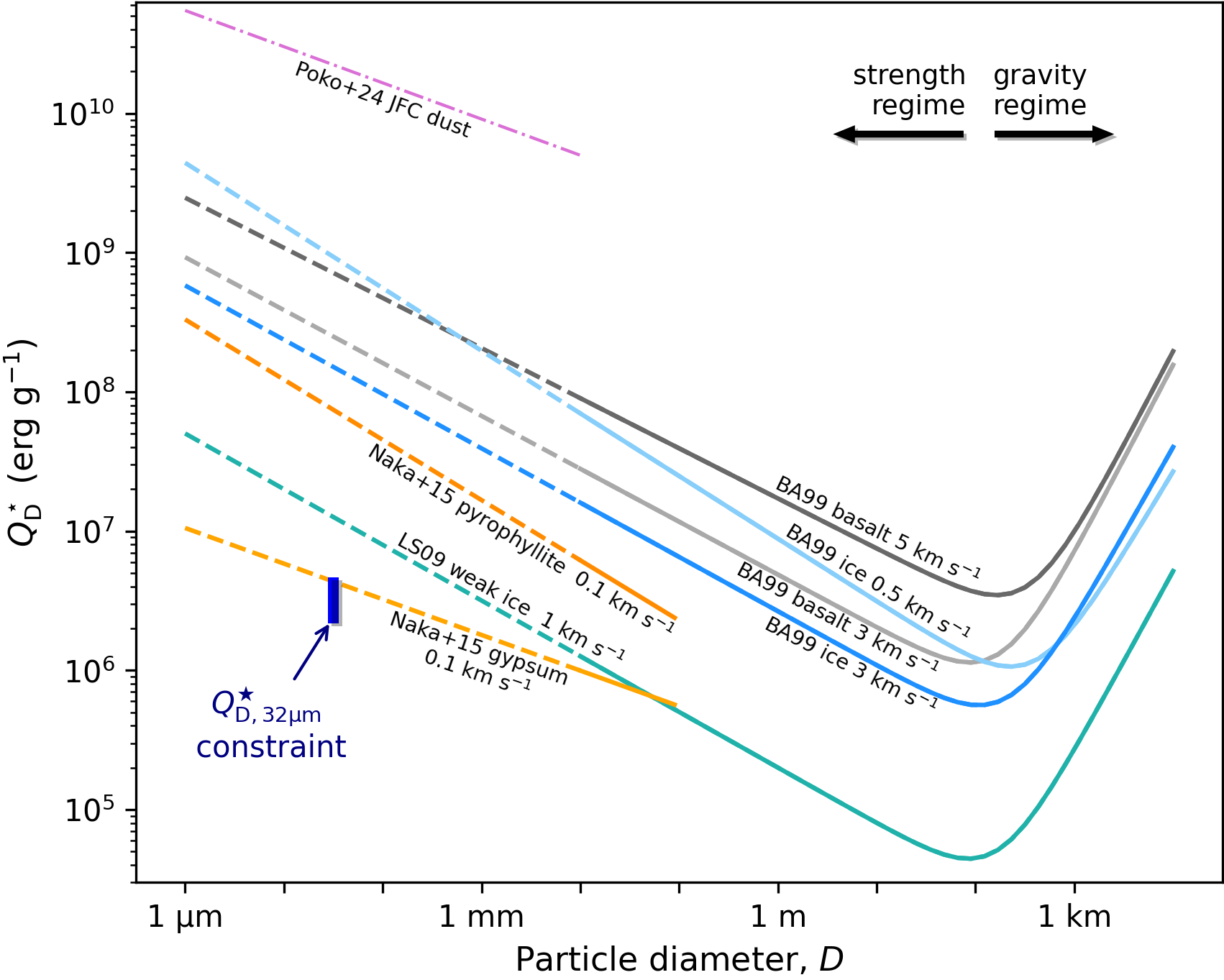}
    \vspace{-3mm}
    \caption{Threshold for catastrophic disruption over particle size 
    for various materials and velocities, as reported by 
    \citet{Benz1999catastrophic, Leinhardt2009full, Nakamura2015size}, 
    shown alongside the constraint from our model fitting.
    These \qdstar{} laws are plotted solid at sizes \qty{>1}{\cm}, the regime
    where these relations are founded in experiments or numerical simulation,
    and dashed at the extrapolated smaller sizes.
    Also depicted is the preferred \qdstar{} prescription by \citet{Pokorny2024how} (dash-dot line),
    constrained by fitting JFC dust models to solar system observables.\vspace{-1mm}}
    \label{fig:qdstar}
\end{figure}

Our model fitting strongly selects for the dust particles' collisional strength parameters, 
\qdnorm{} and \qdslope{}, which determine the size-dependent critical dispersion energy:
$\qdstar = \qdnorm ( D / \unit{\cm} ) ^ {\qdslope}$.
The correlation between these parameters revealed that it is the resulting \qdstar{}
at a few tens of micrometers, rather than the individual parameters, 
that permits a well-fitting model outcome, thus constrained to a range of 
\mbox{(\numrange{2}{4})$\,\times\qty{e6}{\erg\per\g}$}.
We assess this specific constraint by comparing it with extrapolations from conventional
\qdstar{} prescriptions, as illustrated in \cref{fig:qdstar}. 
These prescriptions are based on numerical and experimental simulations
at target sizes greater than \qty{1}{\cm}, depicted with solid lines, 
and are commonly extrapolated to smaller sizes, indicated by dashed lines.

For instance, we include models by \citet{Benz1999catastrophic} for both solid water ice and basalt,
derived from smoothed particle hydrodynamics (SPH) simulations across a broad range of target sizes
(few \unit{\cm} to 100s of \unit{\km}).
These models exhibit the well-known features of $\qdstar (D)$, 
that is, a strength regime with a negative slope for targets sizes \qty{<100}{\m},
where gravitational forces are negligible, and a gravity regime with a positive slope
for larger sizes, where the energy required for dispersal rises steeply 
due to the need to overcome gravitational binding.
\citet{Benz1999catastrophic} found similar magnitudes of strength for basalt and ice,
though basalt targets appeared easier to disrupt at slower impact velocities, 
while the opposite was true for ice. 
However, the ice strengths reported by \citet{Benz1999catastrophic} are considered questionable,
as laboratory experiments have indicated strengths up to two orders of magnitude lower 
\citep{Arakawa1999collisional, Ryan1999laboratory, Giblin2004collisional}.
This discrepancy has prompted the adoption of downscaled `weak ice' prescriptions 
that are more in line with experimental outcomes
\citep{Wyatt2002collisional,Leinhardt2009full}.

In comparison, our constraint on \qdstarcon{} sits slightly below the extrapolations for `weak ice'. 
It is important to note that the collisional velocity for obtaining this constraint
(\qty{127}{\m\per\s}, following our assumptions for the planetesimal belt, see \cref{Sect:ApplFom}),
is smaller than those informing the depicted weak ice strength laws.
However, no definitive trends with velocity were identified across laboratory 
impact experiments with ice within the \qtyrange{100}{1000}{\m\per\s} range, 
potentially due to variations in target preparation \citep{Leinhardt2009full}.

For rocky targets, the speed-dependence of \qdstar{} is better understood.
Based on experimental data covering impact velocities from around \qty{100}{\m\per\s} to
\qty{3}{\km\per\s}, \citet{Nakamura2015size} derived \qdstar{} scaling laws
for porous sulfate (gypsum, porosity \qty{\sim67}{\percent}) and non-porous silicate
(pyrophyllite) dependent on target diameter as well as impact velocity.
Their results, consistent with those for basalt by \citet{Benz1999catastrophic}, 
indicate that the disruption threshold scales with impact velocity.
Notably, this velocity dependence is more pronounced for porous targets
where \qdstar{} increases by a factor of \num{\sim6} when the velocity increases tenfold,
compared to the non-porous targets, where \qdstar{} merely doubles for the same velocity change.
The scaling laws provided by \citet{Nakamura2015size} offer a $\qdstar{}(D)$ curve 
for these materials at velocities akin to those in our model, as illustrated in \cref{fig:qdstar}.
The strength model for porous sulfate aligns with our \qdstarcon{} constraint, 
whereas the model for non-porous rock is about an order of magnitude higher.

While the order-of-magnitude agreement of our \qdstarcon{} constraint with the 
\qdstar{} power laws for `weak ice' and porous rocky material is reassuring,
the validity of extrapolating these power laws from their tested regime at centimetre sizes 
down by \numrange{2}{3} orders of magnitude to tens of micrometers remains questionable.
The possibility of an inflection point and even a positive slope regime at those small sizes
(which in fact provides the best fit of our model) cannot be ruled out.

Moreover, the strength constraints obtained from our model are influenced by several factors. 
Since the analytical model primarily constrains the particles' collisional lifetime, which
depends on collisional velocity, adjustments to model parameters can significantly alter outcomes.
For instance, reducing the semi-opening angle $\epsilon$ by a factor of 1/\nth{3}, that is,
from \qty{1.5}{\degree} to \qty{1.0}{\degree} \citep[as found by][]{Boley2012constraining}, 
requires nearly halving the \qdstar{} normalization to maintain the same optical depth model outcome.
Conversely, an increase in $\epsilon$ by 1/\nth{3} to \qty{2.0}{\degree} 
necessitates nearly doubling \qdstar{}.
An increase of the collisional velocity might also result from considering the 
belt eccentricity or a higher assumed mean proper eccentricity, $e_\mathrm{p}$,
which will similarly raise our constraint on \qdstarcon{}.
Furthermore, cratering collisions, which our current model omits, 
could significantly reduce the collisional lifetimes of particles.
If we assume a universal reduction in collisional lifetimes due to cratering by a factor of five---a
value in line with theoretical considerations \citep{Kobayashi2010fragmentation}---\qdstar{} 
would need to be increased by a factor of \num{\sim7} to maintain equivalent model outcomes.
Such an adjustment would position our \qdstarcon{} constraint in between the 
plotted `weak ice' and pyrophyllite models.

Beyond the more conventional \qdstar{} prescriptions derived from impact experiments,
our constraint can also be compared to the extreme collisional strengths recently deduced by
\citet{Pokorny2024how}, who fitted models of Jupiter-family comet (JFC) micrometeoroids
to solar system observables such as the orbital element distributions of meteors.
Similar previous studies also suggest that cometary meteoroids possess collisional lifetimes
\numrange{1}{2} orders of magnitude longer than those allowed by conventional collisional strengths
\citep{Nesvorny2011dynamical,Pokorny2014dynamical,Soja2019imem2}.
Given that our \qdstarcon{} constraint aligns with the lower end of conventional prescriptions,
it starkly contrasts with these extreme solar system strengths.

\begin{figure} 
    \centering
    \includegraphics[width=\columnwidth,trim={0 0 0 0},clip]{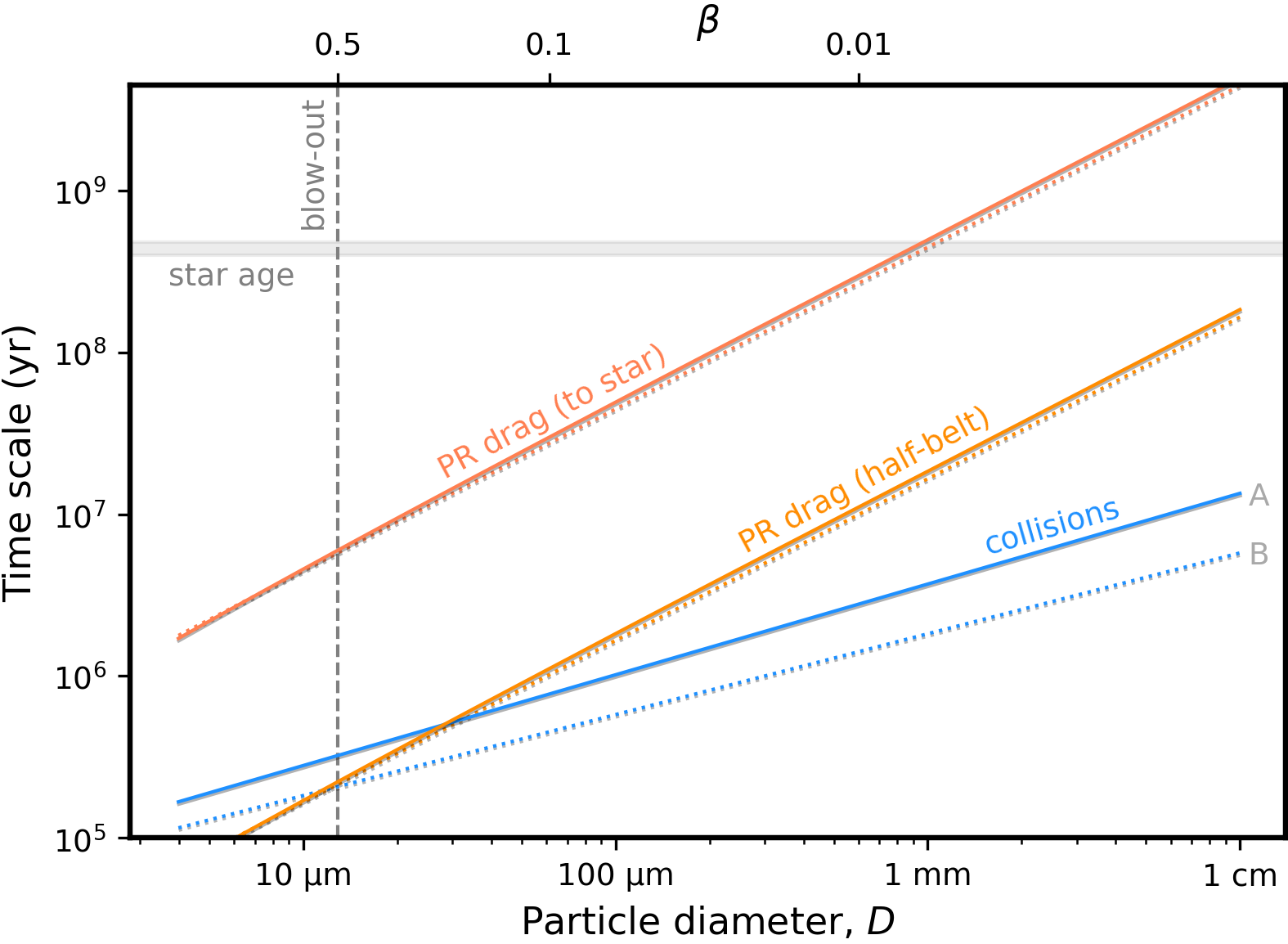}
    \caption{Collisional and PR drag timescales for particles in the planetesimal belt
    in our best-fit model outcomes. 
    Two PR drag timescales are depicted: 
    the time it takes for a (circular) particle orbit to decay from the belt to the star, and
    the time it takes for a (circular) particle orbit to decay from the centre of the belt to its inner edge.
    Scenarios~A and B are shown in solid and dotted lines, respectively.
    Indicated at the top are the $\beta$-factors corresponding to the particle sizes for the material of Scn.~A.
    $\beta$-factors for Scn.~B are marginally larger, with a blow-out size shifted from
    \qty{13}{\um} to \qty{15}{\um}.
    Also indicated is the estimated stellar age of Fomalhaut 
    of \qty{440\pm40}{\mega\year} \citep{Mamajek2012age}.
    }
    \label{fig:lifetimes}
\end{figure}

For reference, \cref{fig:lifetimes} shows the modelled collisional and PR drag time scales
for particles in the Fomalhaut belt, as a function of particle size, for our best-fit outcomes,
which are broadly consistent with the time scales derived by \citet{Su2024imaging} 
for the Vega debris disk---a system with remarkably similar properties to Fomalhaut.

One potential explanation for the discrepancy between the 
conventional strengths and those inferred from solar system observations
is the different collisional velocity regimes being probed.
Collisional velocities in the Fomalhaut belt are on the order of \qty{0.1}{\km\per\s}, 
while grains on JFC orbits colliding in the inner solar system may easily reach relative velocities
of 10s of \unit{\km\per\s} \citep[see typical velocity distributions of JFC meteoroids
at \qty{1}{\astronomicalunit}, \eg][]{Nesvorny2010cometary,Nesvorny2011dynamical,Szalay2019impacta}.
Higher impact velocities and material porosity tend to increase collisional strengths 
for rocky materials \citep{Benz1999catastrophic,Jutzi2010fragment,Nakamura2015size}.
However, extrapolating the high-porosity \qdstar{} models of \citet{Nakamura2015size}
(tested up to \qty{3}{\km\per\s}) to \qty{10}{\km\per\s}
results in \qdstar{} values ranging from \qtyrange{e8}{e9}{\erg\per\g}---still
nearly two orders of magnitude lower than those reported by \citet{Pokorny2024how}.

To speculate, at velocities of 10s of \unit{\km\per\s}, 
which exceed the capabilities of impact experiments with light-gas guns, 
collisional strengths might be higher than anticipated, perhaps due to 
largely damage-free penetration of fluffy grains by highly energetic but small projectiles.
Such a hypothesis would be consistent with the fact that extreme collisional strengths
are in particular required to fit the meteor orbit distributions and
the mass flux to Earth \citep[as indicated in][Fig.~4B therein]{Pokorny2024how}---both
arguably governed by larger micrometeoroids, 
which spend more time near their highly eccentric source orbits
and are thus more exposed to higher collisional velocities.
Similarly, in the studies by \citet{Nesvorny2011dynamical} and \citet{Soja2019imem2}, 
only particle sizes above the \qtyrange{0.1}{1}{\mm} range
require lifetimes significantly extended beyond conventional estimates.

Such reasoning might offer a way to reconcile the extreme strengths
required for the fitting of solar system observations 
with those derived from impact experiments, 
which, conversely, appear compatible with observations of debris disks.

\subsection{Grain composition}
The fitting of our model also constrains the parameters of the dust grain model we employed.
The results suggest that the particles are rather compact composites, 
consisting of porous aggregates of silicate-core, carbonaceous-mantle grains, 
with their voids filled with water ice, comprising \num{50}\%--\num{80}\% of the total particle volume.
Such a composition is generally consistent with what is known about the solar system's Kuiper belt,
where densities measured in (binary) smaller trans-Neptunian objects, typically fall below 
\qty{1}{\g\per\cm^3}, implying considerable ice volume fractions \citep{Barucci2023kuiper}.
More specifically, we may consider the composition of JFCs, which are known
to stem from the Kuiper belt \citep{Duncan1988origin,Fraser2024transition}.
Our constraint of the `refractories-to-volatiles' ratio,
represented as $(m_\mathrm{sil}\!+\!m_\mathrm{carb})\,/\,m_\mathrm{ice}$ in our simplified grain model,
of \numrange{.43}{1.6} (Scn.~A) or \numrange{.43}{0.77} (Scn.~B), 
is compatible with that found in the mass loss of comet 67P/Churyumov-Gerasimenko, which was scrutinized
by the \textit{Rosetta} probe \citep[\numrange{.7}{2.3},][]{Biver2019longterm}, and 
falls on the lower end of the ratios inferred for its nucleus, which show significant discrepancy 
(\eg{} \num{>3}, \citeauthor{Fulle2019refractoryice} \citeyear{Fulle2019refractoryice};
\numrange{.5}{1.7}, \citeauthor{Marschall2025refractorytoice} \citeyear{Marschall2025refractorytoice}).
Substantial amounts of water ice have also been inferred
in other debris disks \citep[\eg][]{Chen2008possible,Lebreton2012icy,Lisse2012spitzer}
and protoplanetary disks \citep[\eg][]{Pontoppidan2005ices,Honda2008detection}. 

The relative abundance of carbonaceous material is not strongly selected by our model, 
as long as it is not too dominant
($\qsi>0.2\Leftrightarrow\qcarb=1-\qsi<0.8$).
This is consistent with the expected cosmic abundance of carbonaceous material
relative to silicates, at a ratio of $m_\mathrm{carb}/m_\mathrm{sil}\!\approx\!0.7$
\citep{Li2003modeling}, equivalent to $\qsi\!\approx\!0.42$.

The structural grain model we employ is generally compatible with cosmic
dust particles collected from the stratosphere, likely originating from the Kuiper belt region
(their origin inferred from cosmic ray exposure ages, 
\citeauthor{Keller2022evidence} \citeyear{Keller2022evidence}; 
though see also \citeauthor{Lin2024solar} \citeyear{Lin2024solar},
who explore additional complexities and unresolved questions in track accumulation modelling),
that is, porous aggregates of mineral grains bound by carbonaceous material.
These may have held ices within their cavities initially, which would have naturally been lost
due to their collection not only within the solar system's sublimation zone
but also after atmospheric entry heating.

This leads us to another significant simplification in our model: the assumption 
that a single particle composition applies universally throughout the entire disk. 
However, in a realistic disk, the ice content of particles may change 
with radial location due to sublimation and/or UV photosputtering.
The sublimation of the water ice is expected to occur once particles reach a temperature of 
\qty{\sim110}{\kelvin} \citep{Kobayashi2008dust}, which in the case of Fomalhaut 
corresponds to a radial distance of \qty{\sim30}{\astronomicalunit}, 
depending on particle material and size.
Yet, we observe no abrupt change of brightness in the resolved disk at around that distance, 
which might be expected from changes in the optical properties of particles 
as ices sublimate from aggregate voids.
In particular, we see no evidence for a sublimation pile-up and subsequent depletion
as theorized by \citet{Kobayashi2008dust}.
A distinct sublimation pile-up requires near-circular particle orbits
\citep[$e\!\lesssim\!0.01$,][]{Kobayashi2008dust,Kobayashi2011sublimation}, 
which may explain its absence, given the Fomalhaut disk's eccentric 
intermediate ring feature \citep[$e\!\approx\!0.3$,][]{Gaspar2023spatially}.
Moreover, an optical depth reduction due to ice loss
might not manifest when more realistic particle properties are considered.
While the Kobayashi model assumes sublimating particles to shrink down to the compact refractory remnants,
a grain model akin to ours may instead leave behind highly porous aggregates after losing its ices, 
potentially maintaining the original cross-section.
Discrete dipole approximation simulations suggest that such irregular porous aggregates
are relatively less sensitive to radiation pressure blow-out 
\citep[with carbon aggregates of \num{\sim75}\% porosity exhibiting only twice
the blow-out size of compact carbon grains around A-type stars,][]{Arnold2019effect}, 
which may enable these remnants to remain bound.

Exposed ices may also be removed from particles by stellar light far-UV photosputtering, 
which around A-type stars such as Fomalhaut could be effective even beyond 
\qty{100}{\astronomicalunit} \citep{Grigorieva2007survival}.
Such early ice removal might seem at odds with the optical properties constrained by our model
(and could help explain the absence of sublimation features).
Indeed, the best fits across our models that exclude water ice entirely 
exhibit deficiencies in total emission at around \qtyrange{60}{150}{\um} 
and in the resolved mid-infrared emission of the interior disk compared to the outer belt,
indicating that the presence of ices is necessary to adequately reproduce these observables
with the pure PR drag transport scenario.
However, far-UV photosputtering ($\lambda_\mathrm{photon}\lesssim\qty{200}{\nm}$) 
can remove ices only from the outer layers of rock-ice mixture particles
that are 10s or 100s of \unit{\um} in size, likely leaving their optical properties
at mid-infrared and longer wavelengths largely unaffected.
Thus, it is conceivable that a more detailed grain model, in which ices are removed
only from the outer layers of particles, could achieve fits similar to our current model,
though exploring this goes beyond the scope of this paper.
Explanations along these lines may also be necessary to reconcile previous studies
that likewise invoked icy particles in debris disks around A-type stars
to simultaneously fit excess SEDs and resolved infrared emission 
\citep{Acke2012herschel,Morales2016herschelresolved,Adams2018dust}
with the photosputtering effect.
Future JWST (near-infrared) scattered light observations may provide more definitive insights into the 
water ice content of particles \citep{Kim2019constraining,Kim2024characterization}.

\begin{figure}
    \centering
    \includegraphics[width=\columnwidth,trim={0 0 0 0},clip]{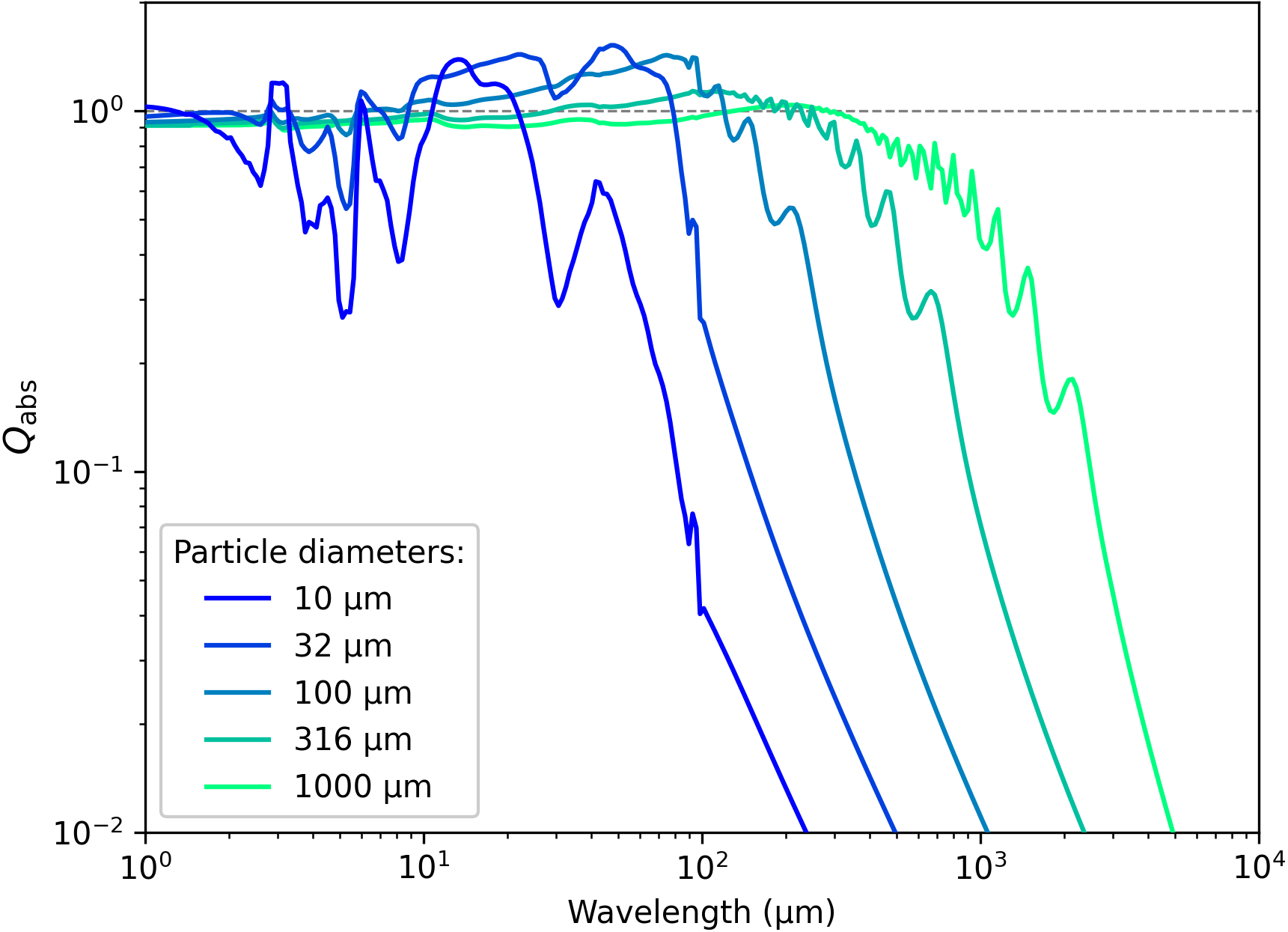}
    \vspace{-4mm}
    \caption{Emissivity profiles ($Q_\mathrm{abs}$) for the best-fit particle model (Scn.~A) 
    as a function of wavelength for various grain sizes.\vspace{-2mm}}
    \label{fig:Qabs}
\end{figure}

For reference, \cref{fig:Qabs} shows the absorption coefficients, $Q_\mathrm{abs}$,
as computed with our optical properties model for our best-fit parameters
(Scn.~A), 
as a function of wavelength and particle size.

\subsection{Planets} \label{sect:planets}
Our model has successfully replicated the observed smooth radial brightness distribution
in the Fomalhaut disk without considering the presence of planets.
Massive planets, however, would be expected to introduce distinct radial features,
such as cavities carved by the ejection of particles from the system in close encounters,
or ring-like overdensities caused by the trapping of particles in exterior mean-motion resonances.
While prominent resonant rings may arguably be suppressed by particle collisions
in a disk as dense as Fomalhaut's \citep{Kuchner2010collisional,Thebault2012planet},
the process of cavity formation through particle ejection should be unaffected.
One may thus be drawn to interpret the absence of clear brightness drops within the extended disk
as an absence of giant planets.

\begin{figure} 
    \centering
    \includegraphics[width=\columnwidth,trim={0 0 0 0},clip]{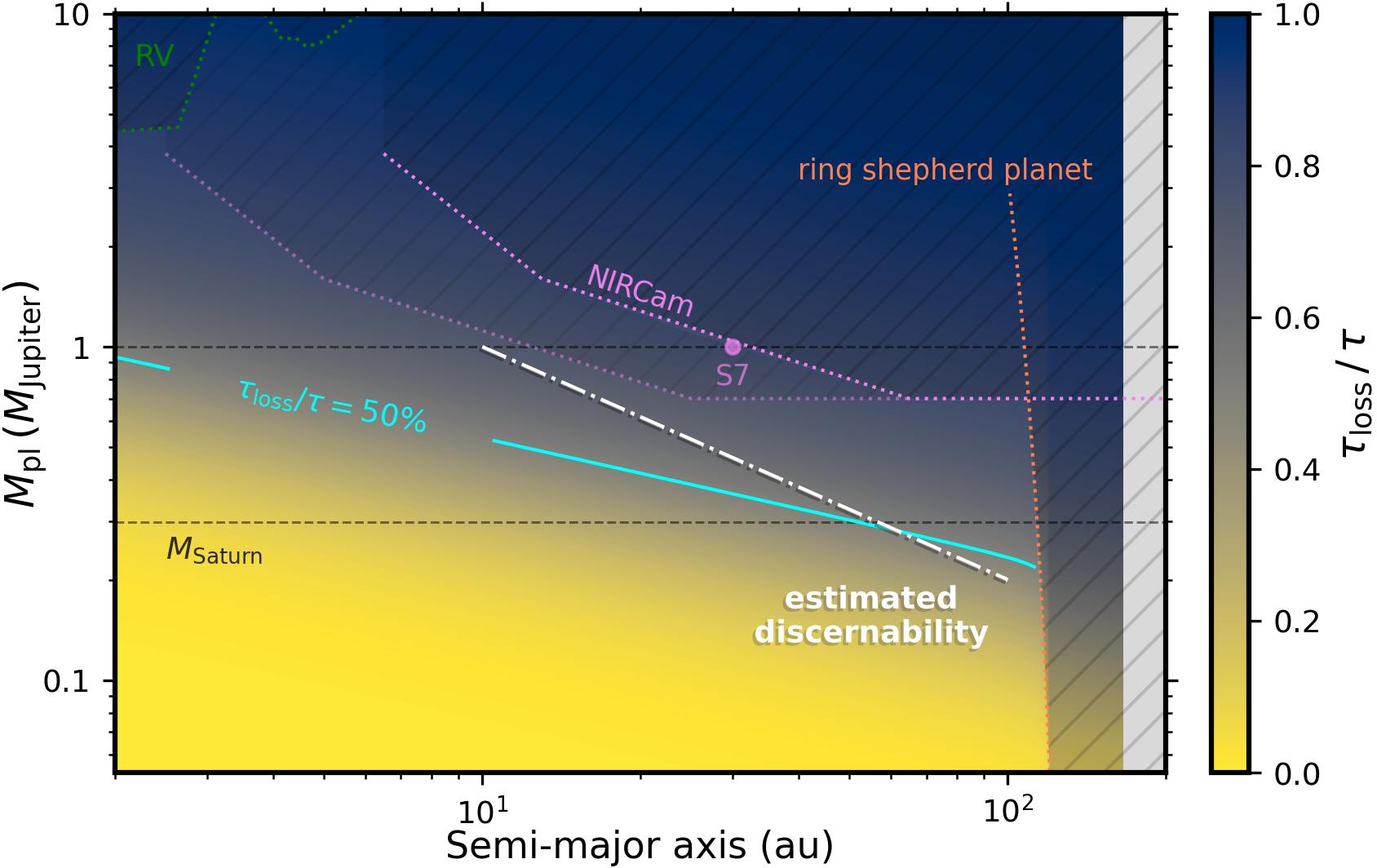}
    \caption{Drop of optical depth due to dust ejection by a hypothetical planet,
    derived for our best-fit model disk with the formalism of \citet{Bonsor2018using}.
    The solid contour (cyan) indicates an optical depth reduction of 50\%.
    Hatched areas bordered by coloured dotted lines represent regimes already probed or ruled out.
    These are informed by the detection limit of a radial velocity survey with HARPS
    \citep[][from their Fig.~7]{Lagrange2013planets},
    the detection limit of the NIRCam planet search 
    \citep[][from their Fig.~7]{Ygouf2023searching}
    (upper and lower violet lines correspond to a planet located on the
    disk minor and disk major axis, respectively),
    as well as the parameters of a hypothetical ring-shepherding planet
    that could have sculpted the inner edge of the planetesimal belt
    \citep{Chiang2009fomalhauts} (planets to the right of this line
    are ruled out as they would have deformed the belt).
    Also indicated are the potential planet parameters of object `S7', 
    revealed by the NIRCam planet search.
    We estimate that planets above the white dash-dot line would have caused
    a discernable optical depth reduction.}
    \label{fig:planets_uplim}
\end{figure}

To explore this further, we apply the formalism developed by \citet{Bonsor2018using},
which calculates the fraction of migrating particles lost to ejection or accretion 
by a planet based on the planet's mass, semi-major axis, and the particles' $\beta$-factor. 
Integrating this model across all particle size bins in our best-fit model disk---weighted
by each bin's contribution to the optical depth at specific radii---enables us
to estimate the total reduction in optical depth for various planetary configurations,
as shown in \cref{fig:planets_uplim}.
This in turn approximates the expected brightness reduction a given planet would introduce.
Generally, the more massive the planet and the larger its semi-major axis, 
the higher the loss in optical depth.

In exploratory tests, we post hoc applied optical depths reductions to our best-fit disk 
according to the \citet{Bonsor2018using} formalism and examined 
the resulting synthetic images and radial brightness profiles.
Crudely, we estimate that optical depth reductions of at least 30\%--70\%,
depending on radial location, would have been discernable,
as indicated by the dash-dot line in \cref{fig:planets_uplim}.
Brightness drops are more readily discerned at larger disk radii 
due to the blurring effect of the PSFs, obscuring smaller-scale structures,
as well as the coronagraphs' transmission drop-off.
Based on this preliminary assessment, we are inclined to rule out the presence
of planets more massive than Jupiter beyond \qty{\sim10}{\astronomicalunit},
and those more massive than Saturn beyond \qty{\sim50}{\astronomicalunit}.

\cref{fig:planets_uplim} also shows the regime recently probed by JWST/NIRCam coronagraphy 
in a search for planets around Fomalhaut, achieving a sensitivity capable of detecting planets
down to approximately Jovian mass \citep{Ygouf2023searching}.
Out of ten detected objects, nine were confirmed as background objects.
Object `S7', however, although detected with only one of the two used filters,
could thus-far not be attributed to a background object.
Follow-up NIRCam observations to confirm or reject this planet candidate,
seemingly located at a separation of \qty{\sim30}{\astronomicalunit} from Fomalhaut, 
are underway \citep{Beichman2023planets}.
Based on our analysis, we would expect a Jovian-mass planet at \qty{30}{\astronomicalunit}
to carve a cavity with an optical depth drop on the order of 80\%.
This is hardly compatible with the observed brightness distribution,
as similarly argued by \citet{Ygouf2023searching}.
Nevertheless, the follow-up observations, which aim for a detection threshold as low as 
\qty{0.3}{\jupitmass} (\ie{} around Saturn's mass), will serve as a test of the 
tentative upper limits we have placed on planetary masses in the Fomalhaut disk.

\citet{Chiang2009fomalhauts} constrained the parameters (\mpl{} and \apl{}) of a hypothetical planet 
that could have sculpted the inner edge of the planetesimal belt as observed in scattered light.
These parameters are represented in \cref{fig:planets_uplim} by an orange dotted line.
The constraints are based on the assumption that the outer boundary of the arising chaotic zone---where
no stable orbits can be maintained---extends just to the inner edge of the planetesimal belt.
This is described by the equation:
\begin{align} \label{eq:chaoticzone}
    a_\mathrm{inner} - \apl \; = \; 2.0 \, \left( \frac{\mpl}{M_\star} \right)^{2/7} \apl  , 
\end{align}
where $a_\mathrm{inner}=\qty{133}{\astronomicalunit}$ is the semi-major axis of the 
inner edge of the planetesimal belt.
\citet{Chiang2009fomalhauts} additionally suggested that the mass of such a ring-shepherding planet
should be \qty{<3}{\jupitmass}, to avoid perturbing the belt particles onto more eccentric orbits,
which would be inconsistent with the belt's observed optical depth profile.

A shepherding planet would also be expected to eject dust migrating inward from the belt,
and thus cause a reduction in optical depth.
Indeed, the MIRI images of the disk indicate a brightness dip occurring near 
the inner edge of the belt, which however our planet-less model reproduces, 
as only a fraction of particles can effectively migrate from the belt 
before being ground down to blow-out sizes.
Nevertheless, the magnitude of this `collisional depletion' depends on the dust's collisional strength,
introducing a degeneracy; the observed dip might not solely result from collisions
but also from planet-induced dynamical depletion.
In our model, adjusting the collisional strength to be up to 10 times greater than in our best-fit
allows us to mimic the observed dip while accounting for the dynamical depletion 
caused by a shepherding planet with a mass up to \qty{.5}{\jupitmass}.
Higher levels of dynamical depletion, however, 
are increasingly difficult to offset with enhanced collisional strength.
This suggests that the pure PR drag scenario for the Fomalhaut disk remains valid
with the inclusion of a belt-shepherding planet up to \qty{.5}{\jupitmass}, 
albeit with collisional strengths somewhat higher than those favoured in the planet-less case.
The inclusion of a shepherding planet of a mass of \qty{.2}{\jupitmass} as hypothesized by
\citet{Janson2020tidal} to explain the transient dust cloud `Fomalhaut~b'
\citep{Kalas2008optical,Galicher2013fomalhaut,Gaspar2020new} via tidal disruption,
necessitates a doubling of collisional strengths to recover an equivalent outcome.
Shepherding planets below \qty{.1}{\jupitmass} cause only marginal dynamical depletion
and do not require an adjustment of collisional strengths to maintain the fit.

It is important to note that the ejection model of \citet{Bonsor2018using} assumes planets 
to be on circular orbits, which may not align with the observed eccentricity 
of the planetesimal belt and the `intermediate ring' brightness feature. 
Interestingly, this `intermediate ring' visible in the F2300C and F2550W images, 
does not appear to have a significant impact on the azimuthally averaged radial brightness profiles. 
This suggests that if a planet is producing this feature (\eg{} through resonant trapping), 
it is not massive enough to eject a substantial fraction of the migrating dust.

\subsection{Implications for other debris disks}
Let us put our findings about the Fomalhaut disk, namely, the plausibility
of a pure PR drag transport scenario, into context.
Before JWST's revelations about Fomalhaut's extensive interior disk,
inferences about its structure typically relied on SED analysis of the unresolved excess,
to determine the underlying dust temperature and thus location.
With that approach, \citet{Su2013asteroid,Su2016inner} derived a warm disk component
with a blackbody temperature of \qty{\sim170}{\kelvin}, thus located
at \qtyrange{\sim8}{15}{\astronomicalunit} and presumably maintained 
by \textit{in situ} dust production by a narrow, collisional asteroid belt analogue.
Therefore, even when considering the degeneracy between the dust location and 
the dust's uncertain emission properties, the discovery of the extended interior disk 
by \citet{Gaspar2023spatially} came as a surprise,
ruling out radially confined \textit{in situ} production as the dust's sole source.

Several 10s of debris disks with unresolved warm components have been identified
whose infrared excesses can be fitted with two-component, 
or rather two-temperature (\ie{} cold and warm) disk models,
suggesting radially confined, warm belts seemingly associated with snowline locations
\citep[\eg][]{Morales2011common,Morales2016herschelresolved,Kennedy2014twotemperature,
Ballering2014probing,Ballering2017what},
thus often interpreted as exo-asteroid belts.
The case of Fomalhaut begs the question:
Is an extended, and thus potentially PR-drag-caused interior disk component
the more common explanation for those unresolved infrared excesses?

In any case, the effectiveness of PR drag transport in explaining the appearance 
of the Fomalhaut disk underscores its relevance in the JWST era. 
Our findings support the conclusions of previous studies, which, 
even before the advent of highly resolved imagery, posited that PR drag alone 
could sustain the inferred abundances of warm dust around specific stars, 
\citep{Reidemeister2011cold,Lohne2012modelling,Schuppler2014collisional,
Mennesson2014constraining,Rigley2020dust}.
Moreover, results on another debris disk captured by JWST/MIRI,
namely that around A-type star Vega, were recently published \citep{Su2024imaging}.
These have likewise revealed a smoothly filled-in disk interior to Vega's planetesimal belt
(though without azimuthal features), consistent with pure PR drag evolution.
More JWST observations specifically investigating the delivery mechanism
of warm dust in debris disk systems are underway.

As established by \citet{Kennedy2015warm},
given the inevitability of PR drag and the relative insensitivity of interior
PR-drag-delivered dust levels to belt density,
all systems with detectable outer planetesimal belts, 
unless hosting a giant dust-ejecting companion, will be significantly permeated with dust.
This has profound implications for the study of these exoplanetary systems.
Notably, dust in habitable zones poses a significant challenge in the search for exo-Earths
\citep[\eg][]{Roberge2012exozodiacal, Currie2023mitigating}, 
and deviations from the nominal (\ie{} PR-drag-delivered) warm dust distribution
can provide insights into the presence of planets or comets 
\citep[\eg][]{Bonsor2018using, Marino2018scattering, Sommer2020effects}.
Recognizing these deviations and understanding their origins will require
precise modelling of the PR drag scenario.
Therefore, assuming our model adequately captures the nature of the Fomalhaut disk,
the particle parameters we have constrained will serve as an essential calibration point
for future efforts aimed at estimating or interpreting the distribution of warm dust
in similar systems.

  \section{Conclusion} \label{Sect:Conclusion}

Our main conclusions are the following:
\begin{itemize}
    \item The observed radial brightness distribution in the Fomalhaut disk is consistent
        with a pure PR drag scenario, that is, no other transport mechanisms appear
        to be necessary to generate the underlying dust distribution.
    \item We find that the faithful replication of instrumental effects and reduction methods 
        is critical when trying to compare model outcomes with disk observables.
        This pertains for instance to the MIRI coronagraphy, 
        where the transmission drop-off at small working angles and PSF blurring 
        considerably skew the derived radial brightness profile of the inclined disk,
        or to \textit{Spitzer} IRS spectrography, where the exact positioning of the 
        extraction field in combination with the applied point-source correction 
        significantly influences the obtained spectrum.
        Taking these effects into account, we find that our model naturally reproduces
        a significant offset between the resolved fluxes in the MIRI F2300C and F2550W images,
        previously noted by \citet[][their Supplementary Section~1.2]{Gaspar2023spatially}, 
        despite more comparable fluxes at these wavelengths in the underlying source and synthesized spectra.
        This suggests that these datasets can be reconciled without necessarily invoking an issue
        with the F2300C reduction pipeline.
    \item We constrain the collisional strength of the particles in this pure PR drag scenario, 
        specifically the catastrophic disruption threshold \qdstar{} at a particle size of 
        $D\!\approx\!\qty{32}{\um}$ to a range of \mbox{(\numrange{2}{4})$\,\times\qty{e6}{\erg\per\g}$}.
        This roughly aligns with conventional strength scaling laws derived from impact experiments
        for solid ice \citep{Leinhardt2009full} as well as porous rock \citep{Nakamura2015size}.
        The extreme strengths recently deduced for solar system cometary grains \citep{Pokorny2024how}
        are contradicted by our findings, though this might be due to the different velocity regimes being probed.
        Accounting for cratering collisions, which our model omits,
        may increase our \qdstar{} constraint by up to an order of magnitude.
    \item We constrain the disk particles to be rather compact composites,
        consisting of porous aggregates of silicate-core, carbonaceous-mantle grains, 
        with their voids filled with water ice, comprising \num{50}\%--\num{80}\% of the total particle volume.
        Notably, our simplified grain model, which does not account for changes in composition,
        effectively reproduces the observed radial brightness distribution,
        which in fact lacks the hypothesized sublimation depletion at the snowline 
        of icy dust disks \citep{Kobayashi2008dust}.
        While we do not suggest that ices survive within the snowline (\qty{\sim30}{\astronomicalunit}),
        the absence of a discernible sublimation-induced depletion in the disk may indicate
        that particle behaviours---such as sublimation from the voids of a porous refractory matrix
        that largely preserves optical cross-section---differ from previous theoretical expectations.
        Further study is necessary to explore the impact of grain structure
        on sublimation dynamics and disk appearance.
    \item The radial (\ie{} azimuthally averaged) brightness distribution of the Fomalhaut disk
        shows no clear indications of disturbances by massive planets.
        Given the expected cavity formations that such bodies would cause
        in a disk primarily sculpted by PR drag, our findings suggest the absence of 
        Jovian-mass planets beyond \qty{10}{\astronomicalunit} and 
        Saturnian-mass planets beyond \qty{50}{\astronomicalunit} within the extended inner disk.
        Nevertheless, an outer-belt-shepherding planet of up to \qty{\sim.5}{\jupitmass} could
        still be compatible with a PR-drag-maintained interior disk, though requiring
        higher collisional strengths to offset the dynamical depletion.
        An upcoming JWST/NIRCam planet search with $\sim$Saturn-mass sensitivity will
        not only test these upper limits, but also, both the detection or non-detection
        of a shepherding planet will help refine the constraints we can place on collisional strengths.
    \item Given the inevitability of PR drag, 
        all systems with detectable outer belts---unless obstructed
        by a dust-ejecting giant planet---are likely permeated with dust.
        Recently, high levels of pervasive inner dust have also been found around Vega
        \citep{Su2024imaging}, supporting this picture.
        More JWST observation programmes focussing on warm dust in debris disks
        are underway \citep{Han2024what,Matra2024multiwavelength},
        which will test the generalizability of our results for Fomalhaut.
        The particle parameters we have constrained,
        particularly the collisional strength at the relevant sizes,
        which thus far relies solely on extrapolations,
        establish a calibration point for modelling PR-drag-delivered dust around other stars,
        which is essential for predicting dust levels in habitable zones
        and for interpreting system architectures from resolved images.
\end{itemize}

\section*{Acknowledgements}
We are grateful for the support by United Kingdom Reasearch and Innovation (UKRI) 
- Science and Technology Facilities Council (STFC) grant no.~ST/W000997/1.
We thank Grant Kennedy for his support and for providing the Fomalhaut photosphere model,
as well as Amy Bonsor and Marija Janković for insightful discussions.
We also thank Kate Su and the reviewer, Petr Pokorný, for their valuable comments and suggestions.
This work is based in part on observations made with the NASA/ESA/CSA James Webb Space Telescope.
The JWST observations are associated with programme 1193.

\phantomsection
\label{Sect:DataAvailability}
\section*{Data Availability} 
The JWST/MIRI data underlying this article are provided by \citet{Gaspar2023spatially}, available at 
\href{https://github.com/merope82/Fomalhaut/}{https://github.com/merope82/Fomalhaut/}.
The core components of our modelling framework, that is, 
our implementation of the PR drag disk model from \citet{Rigley2020dust}---extended
with the capability to generate astrophysical scenes---and 
our method to find dust optical properties and temperatures,
are available as separate \texttt{Python} packages,
\texttt{pyrdragdisk} and \texttt{astrodust\_optprops}, respectively, at
\href{https://github.com/rigilkent/pyrdragdisk}{https://github.com/rigilkent/pyrdragdisk}
and 
\href{https://github.com/rigilkent/astrodust\_optprops}{https://github.com/rigilkent/astrodust\_optprops}.

    \bibliographystyle{mnras}
    \bibliography{references} 
	
  \label{lastpage}
\end{document}